\documentclass[manuscript, screen]{acmart}

\AtBeginDocument{%
  \providecommand\BibTeX{{%
    \normalfont B\kern-0.5em{\scshape i\kern-0.25em b}\kern-0.8em\TeX}}}

\usepackage{slashbox}
\usepackage[T1]{fontenc}% optional T1 font encoding
\usepackage{subfig}
\usepackage{caption}
\usepackage{multirow}
\usepackage{tikz}

\usepackage{xcolor}

\newcommand{\sky}[1]{\textcolor{black}{#1}}
\newcommand{\skyRe}[1]{\textcolor{black}{#1}}

\begin{document} 

%
% The "title" command has an optional parameter, allowing the author to define a "short title" to be used in page headers.
\title{Machine Learning Based Cyber Attacks Targeting on Controlled Information: A Survey}

%
% The "author" command and its associated commands are used to define the authors and their affiliations.
% Of note is the shared affiliation of the first two authors, and the "authornote" and "authornotemark" commands
% used to denote shared contribution to the research.
\author{Yuantian Miao}
\affiliation{%
  \institution{School of Software and Electrical Engineering,
Swinburne University of Technology}
  \streetaddress{Swinburne University of Technology, Hawthorn}
  \city{Melbourne}
  \state{VIC}
  \postcode{3122}}
\email{ymiao@swin.edu.au} 
\author{Chao Chen}
\affiliation{%
  \institution{School of Software and Electrical Engineering,
Swinburne University of Technology}
  \streetaddress{Swinburne University of Technology, Hawthorn}
  \city{Melbourne}
  \state{VIC}
  \postcode{3122}}
\email{chaochen@swin.edu.au}

\author{Lei Pan}
\affiliation{%
 \institution{School of Information Technology,
Deakin University}
 \streetaddress{Deakin University, Geelong}
 \city{Melbourne}
  \state{VIC}
  \postcode{3220}}
\email{l.pan@deakin.edu.au}
 
\author{Qing-Long Han}
\affiliation{%
  \institution{School of Software and Electrical Engineering,
Swinburne University of Technology}
  \streetaddress{Swinburne University of Technology, Hawthorn}
  \city{Melbourne}
  \state{VIC}
  \postcode{3122}}
\email{qhan@swin.edu.au}

\author{Jun Zhang}
\affiliation{%
  \institution{(corresponding author) School of Software and Electrical Engineering,
Swinburne University of Technology}
  \streetaddress{Swinburne University of Technology, Hawthorn}
  \city{Melbourne}
  \state{VIC}
  \postcode{3122}}
\email{junzhang@swin.edu.au}

\author{Yang Xiang}
\affiliation{%
  \institution{School of Software and Electrical Engineering,
Swinburne University of Technology}
  \streetaddress{Swinburne University of Technology, Hawthorn}
  \city{Melbourne}
  \state{VIC}
  \postcode{3122}}
\email{yxiang@swin.edu.au}

%
% By default, the full list of authors will be used in the page headers. Often, this list is too long, and will overlap
% other information printed in the page headers. This command allows the author to define a more concise list
% of authors' names for this purpose.
\renewcommand{\shortauthors}{Miao et al.}

%
% The abstract is a short summary of the work to be presented in the article.
\begin{abstract}
Stealing attack against controlled information, along with the increasing number of information leakage incidents, has become an emerging cyber security threat in recent years. Due to the booming development and deployment of advanced analytics solutions, novel stealing attacks utilize machine learning (ML) algorithms to achieve high success rate and cause a lot of damage. Detecting and defending against such attacks is challenging and urgent so that governments, organizations, and individuals should attach great importance to the ML-based stealing attacks. This survey presents the recent advances in this new type of attack and corresponding countermeasures. The ML-based stealing attack is reviewed in perspectives of three categories of targeted controlled information, including controlled user activities, controlled ML model-related information, and controlled authentication information. Recent publications are summarized to generalize an overarching attack methodology and to derive the limitations and future directions of ML-based stealing attacks. Furthermore, countermeasures are proposed towards developing effective protections from three aspects---detection, disruption, and isolation.
\end{abstract}


%
% Keywords. The author(s) should pick words that accurately describe the work being
% presented. Separate the keywords with commas.

%
% This command processes the author and affiliation and title information and builds
% the first part of the formatted document.
\maketitle

\section{Introduction} \label{sec_intro}
% Background: information leakage -> serious problem -> examples
Driven by the needs to protect the enormous value within data and the evolution of the emerging data mining techniques, information leakage becomes a growing concern for governments, organizations and individuals \cite{alneyadi2016survey}. Compromising the confidentiality of protected information is an information leakage incident and a prominent threat of cyber security \cite{ahmadian2018information}, for instance, the leakage of sensitive information results in both financial and reputational damages to the organizations \cite{cheng2017enterprise}. Thus, information leakage incidents are indeed an urgent threat that deserves the public attention.

% information leakage in cybersecurity -> stealing controlled information -> controlled information definition with categories (definition: focus on confidentiality) -> attack examples 
This survey introduces the stealing attack in the cyber security area. According to \cite{flavian2006consumer}, the information leakage can be defined as the violation of confidentiality of methods/mechanisms/framework which stores information or has access to information. In other words, the introduced attack aims at stealing the controlled information. According to the cyber attack definition in \cite{kissel2013glossary}, the term ``controlled'' has an implicit meaning as ``protected''. Comparing to the attack compromising of a computing environment/infrastructure or data integrity, it is more difficult to detect the controlled information stealing attack in advance. Cyber attacks that is defined as ``disrupt, disable, destroy, or maliciously control a computing environment/infrastructure and destroy the integrity of the data'' \cite{kissel2013glossary} are out of the scope of this survey, for example, a DDoS attack leaking customer data \cite{califano2015using} is excluded. According to the literature collected between 2014 and 2019, there are three common vulnerabilities subject to the controlled information stealing attacks: 
\begin{enumerate} 
 \item \emph{User activity information} is a primary target, especially the one stored on the mobile devices. For example, \cite{diao2016no} extracted user's foreground app running in Android in order to exploit it for the phishing attack, while the user activity information was protected by a nonpublic system level permission \cite{OgunwaleAndroid}. 
 \item \emph{ML models and their training data} are also exploited, particularly those which are hosted on the Machine-Learning-as-a-Service (MLaaS) systems. For instance, an ML model is confidential due to the pay-per-query development in a cloud-based ML service \cite{tramer2016stealing} as well as the security mechanisms contained in spam/fraud detection applications \cite{biggio2013evasion,huang2011adversarial, lowd2005adversarial,laskov2014practical}.
 \item \emph{Authentication information} is the third category such as keystroke information, secret keys, and passwords.
\end{enumerate}

% machine learning in stealing controlled information attack (exclude malware)
As a fast-growing technique in the recent years, ML techniques are applied widely in various cyber security areas, such as cyber attack prediction \cite{sun2019data}, insider threat detection \cite{liu2018detecting}, network traffic classification \cite{zhang2015robust,zhang2013network,zhang2013internet,liu2017fuzzy}, spam detection \cite{chen2017statistical}, and software vulnerability detection \cite{lin2018cross}. MLaaS \cite{ribeiro2015mlaas} assists users with limited computing resources or limited ML knowledge to utilize ML models. \cite{zhang2018level} leveraged the ML framework to enhance the accuracy of a user activities inference attack. Because of the effectiveness and efficiency brought by ML techniques to the stealing attack, the loss resulted from the ML-based stealing attack is significant. In this survey, the ML-based stealing attack is defined as follows: \emph{An attacker utilizes an ML algorithm to build up a computational model to disclose the controlled information, while the raw dataset is collected in the legitimate ways}. This definition is explained in two attack modes according to the targets: In the first attack mode, attackers aiming to perform an accurate and efficient stealing attack
build up an ML model as a tool to  derive the targeted controlled information; in the second attack mode, the ML model itself is the target. Building up the ML model means reconstructing the targeted controlled information --- the model within an MLaaS platform, which is also known as model reconstruction attack \cite{fredrikson2015model}. Two types of the ML-based stealing attacks are summarized in this survey. But other attacks, which leak controlled information without applying ML techniques, have been surveyed in \cite{ferrag2017privacy} and \cite{barona2017survey}. Furthermore, malware was introduced to leak password files in \cite{guri2018bridgeware}, and an eavesdropping attack was proposed in \cite{zeng2016active} to increase the information leakage rate without using ML algorithms. This survey investigates the ML-based stealing attack.

\begin{figure*}[!ht] %% Paper hierarchical 
\centering
\includegraphics[width=.55\textwidth]{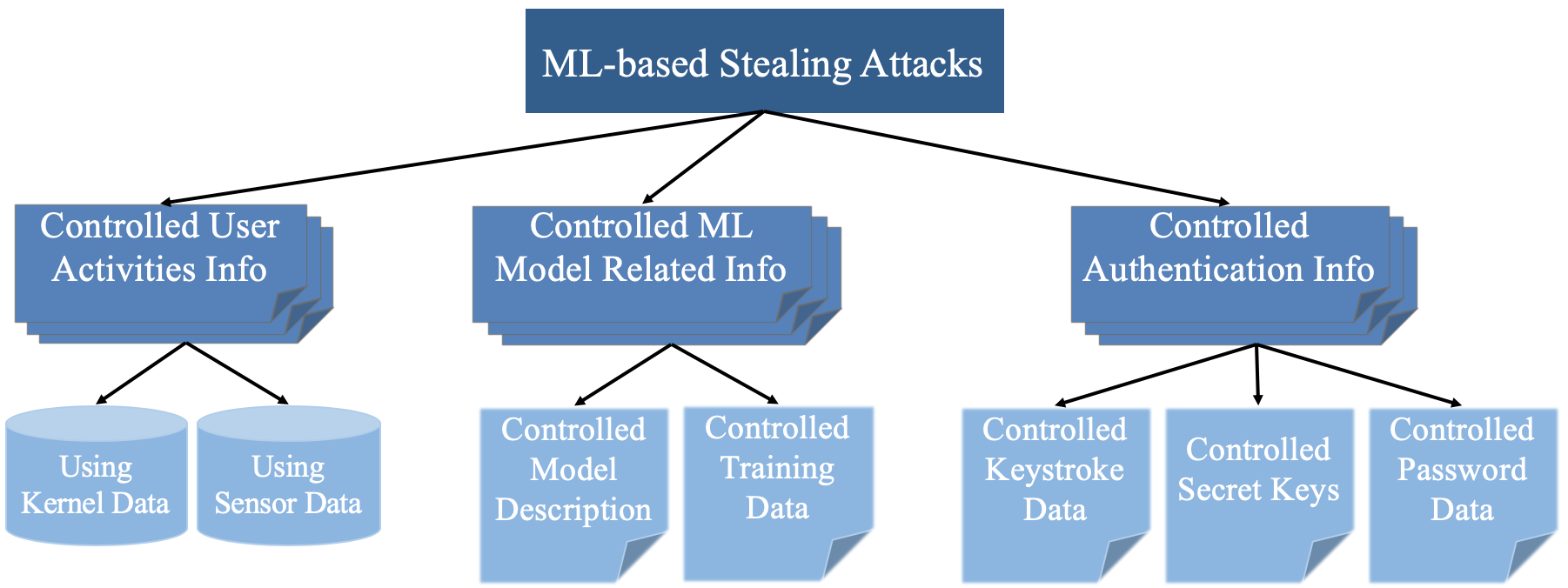}
\caption{Introduced Stealing Controlled Information Attack Categories. (Info: information)} 
\label{fig:intro_categories}
\end{figure*}

% compare to similar survey (data leakage, information leakage, nan's survey of data breach incident, liu's insider attack survey -> difference)
There are a few surveys related to information leakage/data breach illustrating the threat and/or prevention. The prevention of leaking  confidential data was studied in \cite{alneyadi2016survey} through security procedures (i.e.~information security policies) and regular security mechanisms (e.g.~intrusion detection system). The conducted leakage is mainly caused by the incomplete predefined rules of these security mechanisms. This survey analyzes the information leakage problem from the viewpoint of insufficient predefined rules (i.e.~permissions) and exploitable weaknesses of the target infrastructure. In addition, the threat of data breach towards cloud computing was reviewed in \cite{barona2017survey} with primary causes to data breaches as malicious codes, hacked system, electronic back-ups, malicious insiders, and lost devices. A similar conclusion was drawn in \cite{ferrag2017privacy} in the context of ad hoc social network. The attacks of leaking privacy was summarized in \cite{ferrag2017privacy} including eavesdropping-based attack, man-in-the-middle attack, DoS attack, and so on. This survey investigates the ML-based attack using ML techniques to steal information in a legitimate way instead of breaking or hacking the security mechanisms.

A concept for data leakage prevention and information leakage prevention is proposed in \cite{hauer2015data}, that is, the previous attacks were categorized by the causes of leakage prior to the implementation of the prevention. Furthermore, similar analysis results of the attack categories were found in \cite{barona2017survey}. The protection methods will be reviewed based on the analysis of identified information leakage, but the explored attacks are categorized by the types of the controlled information. In \cite{liu2018detecting}, malicious insiders could steal confidential information if the data is shared within the entire user hierarchy. This survey conducts the ML-based stealing attack from both outsiders' and insiders' (i.e.~as a participant in the GAN attack) viewpoints.

% Collection & Structure
The core papers reviewed in this survey were primarily selected from the top four conferences in cyber security research field from 2014 to 2019, which are, IEEE S\&P, ACM CCS, NDSS, and USENIX Security Symposium. The four keywords used in our search are ``\textbf{leak}'', ``\textbf{information}'', ``\textbf{train}'' and ``\textbf{attack}''. Then we further filtered out the papers without the keyword ``machine learning''. We refined the paper collection based on the citation counts. That is, by checking the papers published in other high quality conferences and journals, we selected the papers with high citations as the core papers.

% Contribution % reorganize: identification & classification; review & summarize methodology; attack challenge summarize & future direction & protection direction
This survey introduces a new rising threat of stealing controlled information, and catches up with the trends of this kind of stealing attack and its countermeasures. Our contributions can be itemized as follows:
\begin{itemize}
\item We introduce the ML-based stealing attack, which aims at stealing the controlled/protected information and leads to huge economic loss. Herein, ML algorithms are applied in the attack to increase the success rate in various aspects. The classification of the ML-based stealing attacks is built based on the targeted controlled information preferentially. Based on this classification, the vulnerabilities in various systems and corresponding attacks are sorted out and revealed.
%The knowledge of classification indicates the vulnerability of various security guarantees for different controlled information.
\item We survey the advances of the ML-based stealing attacks between 2014 and 2019. A methodology applied for the ML-based stealing attack against the controlled information is generalized to five phases --- reconnaissance, data collection, feature engineering, attacking the objective, and evaluation. The methodology highlights the similarity of these attacks from strategies and technical perspectives. The public datasets used for the attack analysis are also summarized and referenced correspondingly. 
\item We discuss the challenges of attacks stealing controlled information and forecast their future directions accordingly. Since controlled information security is a subject of competition between attackers and defenders, the countermeasures are summarized and discussed based on the analysis of the ML-based stealing attacks.
\end{itemize}
By improving our knowledge of this emerging attack, the ultimate purpose of this survey is to safeguard the information thoroughly. In the information era, the leakage of information, especially those have already been controlled, will result in tremendous damage to both corporations and individuals \cite{cheng2017enterprise,InfoWatch,IBM,JuniperResearch}. This survey reveals that current protections cannot fully suppress the existing ML-based stealing attacks. As discussed in Section \ref{sec_discussion}, in the near future, protecting controlled information can be improved from detecting the access states of related data, disrupting the related data with considerable utility, and isolating the related data from being accessed.

% Structure
The rest of this survey is organized as follows: The stealing attack methodology is summarized in Section \ref{sec_method}. In Section \ref{sec_LR}, the literature review of the stealing attack using ML algorithms in the past five years is presented, where stealing attacks are reviewed in three categories classified by the types of targeted controlled information. In Section \ref{sec_discussion}, challenges of the attack are discussed and followed by corresponding future directions. Section \ref{sec_conclusion} concludes the survey.

\section{Attack Methodology: MLBSA} \label{sec_method}
%%%%% Methodology: Define Targeted Controlled Information; Analyze Accessible Data to Perform Attack; Machine Learning-based Attack; Attack Evaluation %%%%%%% define a few concepts about steps for each phases.

\begin{figure}[!th] %% Methodology workflow %% 
\centering
\includegraphics[width=.3\textwidth]{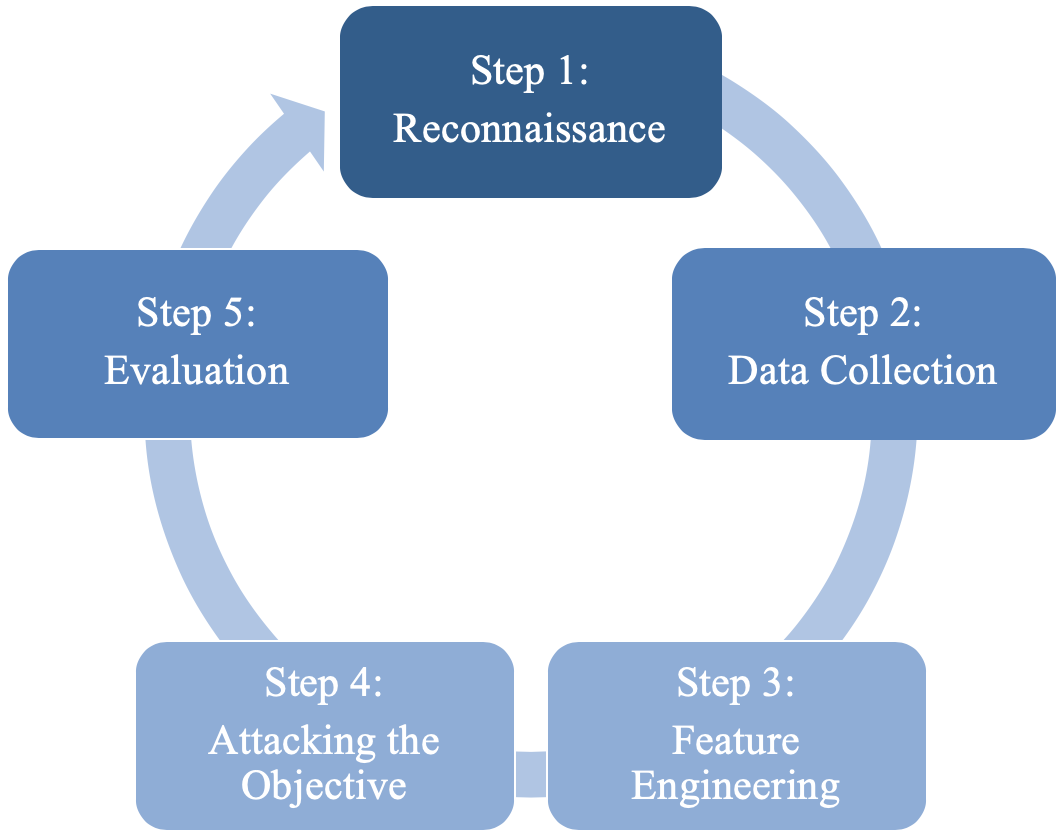}
\caption{ML-based stealing attack methodology (abbreviated as MLBSA methodology).} 
\label{fig:method_wf}
\end{figure}

%Cybersecurity is a game between attackers and defenders. 
This section proposes an attack methodology for stealing controlled information attacks utilizing ML techniques as shown in Fig.~\ref{fig:method_wf}. And the methodology is named as the Machine Learning Based Stealing Attack (MLBSA) methodology. We revised the cyber kill chain \cite{khan2018cognitive,sun2019data} for modeling the MLBSA methodology. A typical kill chain consists of seven stages including reconnaissance, weaponization, delivery, exploitation, installation, command and control, and actions on objectives \cite{yadav2015technical,kiwia2018cyber}. \textit{Reconnaissance} aims to identify the target by assessing the environment. As a result, the prior knowledge of attacks can guide data collection. Regarding the ML-based stealing attack, \textit{weaponization} means data collection. Extracting the useful information via feature engineering is essential. Using supervised learning, the ML-based model is built as a weapon taking \textit{actions on objectives}. Moreover, the ML-based stealing attack may keep improving its performance and accumulate the knowledge gained from its retrieved results. Other stages of kill chain, including \textit{delivering} the weapon to the victim, \textit{exploiting} the vulnerabilities, \textit{installing} the malware, and using \textit{command} channels for remote \textit{control} \cite{kiwia2018cyber}, are considered as a preparation phase before attacking the objectives. In this paper, the preparation phase is named feature engineering. Having consolidated a few steps of the kill chain, the MLBSA methodology consists of five phases, which are organized in a circular form implying a continuous and incremental process. The five phases of the MLBSA methodology are 1) reconnaissance, 2) data collection, 3) feature engineering, 4) attacking the objective, and 5) evaluation. The following subsections will illustrate each phase in details.

\subsection{Reconnaissance} \label{sec_method_reconnaissance} %%%%%%% define targets & analyze accessible data %%%%%% 
Reconnaissance refers to a preliminary inspection of the stealing attack. The two aims of this inspection include defining adversaries' targets and analyzing the accessible data in order to facilitate the forthcoming attacks. 

The target of adversaries in the published literature is usually the confidential information controlled by systems and online services. According to Kissel \cite{kissel2013glossary} and Dukes \cite{dukes2015committee}, the term ``information'' is defined as ``the facts and ideas which can be represented as various forms of data, within which the knowledge in any medium or form are communicated between system entities''. For example, an ML model (e.g.~prediction model) represents the knowledge of the whole training dataset and can act as a service to return results of any search queries \cite{wang2018stealing,tramer2016stealing}. Thereby, the controlled information can be interpreted as the information stored, processed, and communicated in the controlled area for which the organization or individuals have confidence that their protections are sufficient to secure confidentiality. It is more difficult to detect the attack against confidentiality than the attacks against integrity and availability. These information stealing attacks are often referred to as the ``unknown unknowns''. 

In this survey, the targeted controlled information can be classified into three categories: user activities information, ML related information, and authentication information. Herein, user activities of the mobile system, such as which app is running in the foreground, are considered as sensitive information. Such sensitive information should be protected against security threats like phishing \cite{diao2016no}. ML models can be provided on the Internet as a service to analyze the big data and build predictive models \cite{wang2018stealing, tramer2016stealing}, such as Google Prediction API and Amazon SageMaker. In this scenario, both the model and training data are considered to be confidential subjects. However, when some ML services allow white-box access from users, only training data is considered as confidential information. For example, passwords and secret keys unlocking mobile devices and authenticating online services should always be stored securely \cite{gras2018translation,wang2016targeted,melicher2016fast, veras2014semantic}. Using ML to infer the password from user's keystrokes breaks the information confidentiality \cite{sun2016visible, liu2015good}.

Since the information that adversaries aimed to steal is in control, the accessible data is the breakthrough point. In order to analyze its value, the attacker mimics a legitimate user to learn the characteristics and capabilities of the targeted systems, especially those related to the controlled information. During the reconnaissance of accessible data, the attacker needs to exhaustively search all possible entry points of the targeted system, reachable data paths, and readable data \cite{spreitzer2018procharvester}. When the attacker aims at user's activities, the triggered hardware devices and their corresponding logged information will be investigated \cite{diao2016no, spreitzer2018procharvester, zhang2018level}. For example, the attacker always searches and explores the readable system files, such as interrupt timing data \cite{diao2016no, spreitzer2018procharvester} and network resources \cite{zhang2018level}. To perform a successful stealing attack against a model \cite{wang2018stealing, tramer2016stealing} or training samples \cite{shokri2017membership, fredrikson2015model, hitaj2017deep}, the functionalities of ML services (e.g.~Amazon ML) are analyzed by querying specific inputs. The attacker analyzes the relationship between the inputs and the outputs including output labels, their corresponding probabilities (also known as confidence values), and the top ranked information \cite{fredrikson2015model, shokri2017membership, tramer2016stealing}. The relationship reveals some internal information about the target model and/or training samples. For authentication information stealing attacks, stealing keystroke information needs to utilize some sensor activity information activated by the attacker, while the intermediate data can be regarded as accessible data \cite{sun2016visible,liu2015good}. The fine-grained information about security domains, e.g.~secret keys, can be inferred by analyzing the accessible cache data \cite{gras2018translation, zhou2016software}. The accessible data related to the target information are defined in the reconnaissance phase.

\subsection{Data Collection} \label{sec_method_collection} %%%%%%%% query design & query content %%%%%%%% 
Having conducted a detailed reconnaissance process, the attacker refines the scope of their targeted controlled information alongside with the awareness of the related accessible data. Then the attacker designs specific queries against the target system/service to collect useful accessible data. What differentiates the information stealing attack and other forms of cyber attacks is that the datasets are collected in a malicious manner instead of being collected via the malicious way. In accordance with the intelligence gained during the reconnaissance phase, data collection can be either active collection or passive collection.

Active collection refers to that the attacker actively interacts with the targeted system for data collection. Specifically, an attacker designs some initial queries to interact with the system and subsequently collects the data. The goal of the attacker guides the design of malicious interactions, referring to the analysis results from the reconnaissance phases. All data closely related to the controlled information can be gathered as a dataset for the stealing attack. For example, if an attacker intends to identify which app is launched in a user's mobile, some system files like $procfs$ recording app launching activities should be collected, according to \cite{spreitzer2018procharvester}. In addition, various apps will be launched for several times by the attacker for app identification. By running 100 apps, a dataset of kernel information about the foreground UI refreshing process was gathered in \cite{diao2016no} to identify different apps. The active collection for stealing keystroke information and secret keys are similar to that for stealing user activities information. Interacting with the operating system with different keystrokes inputs, sensor information like acceleration data and video records was collected in \cite{sun2016visible, liu2015good} to infer keystroke movements. Moreover, cache data was collected in \cite{gras2018translation, zhou2016software} to analyze the relationship between memory access activities and different secret keys.

Additionally, the active collection for stealing ML-related information aims to design effective and efficient queries against the target model. The design of active collection varies when the target model allows black-box access or white-box access (see more details in Section~\ref{sec_LR}). With the black-box access, the inputs and corresponding outputs are collected to reveal the model's internal information. The number of inputs should be sufficient to measure the model's functionality \cite{tramer2016stealing, wang2018stealing} or clarify its decision boundary \cite{papernot2017practical,joon2018towards,fredrikson2015model}.  A set of inputs was synthesized in  \cite{papernot2017practical} to train a local model to substitute the target model. The range of inputs should also be wide enough to include samples inside and outside a model's training set \cite{shokri2017membership,salem2019ml}. With the white-box access, not only inputs and outputs but also some internal information are collected to infer the training data information. For instance, the model was updated in \cite{hitaj2017deep,melis2019exploiting} by training a local model with data having different features, and the changes in the global model's parameters are utilized to infer the targeted feature values.

The other kind of collection is passive collection, which is defined as gathering all data related to the targeted controlled information without engaging with the targeted system/service directly. The passive collection mainly used to steal the password information in this survey where the targeted system/service is always a login system or permission granted. In such a case, engaging with the target system/service directly could only provide attackers the information on whether the guessing password is correct or not. This information can be used to validate an ML-based attack but is unable to contribute to an attack model as training or testing data.  Users' passwords were cracked in \cite{wang2016targeted} by generating many passwords in high probabilities based on people's behaviors of password creation, while passwords were generated in \cite{veras2014semantic} based on the semantic structure of passwords. Specifically, attackers collect the relevant information such as network data, personal identifiable information (PII), previous leaked passwords, the service site's information and so on \cite{wang2016targeted,veras2014semantic, melicher2016fast}. The information can be gathered by searching online and accessing some open sources data like leaked passwords (as shown in Table~\ref{tab:authen_psw}). The attack targeting at password data primarily uses the passive collection to gather information.

All of the stealing attacks involved in this survey use the supervised learning algorithms. The ground truth need to be established in the data collection phase. Among the investigated ML-based stealing attacks, collected data was labeled with the target information or something related. For instance, the kernel data about the foreground UI refreshing process caused by app launching activities is labeled with corresponding apps \cite{diao2016no}. In \cite{shokri2017membership, salem2019ml}, data was labeled with member or non-member of the target training set. Similarly, for stealing the authentication information, the collected sensor data was labeled in \cite{liu2015good,sun2016visible} with corresponding input keystrokes. The ground truth of datasets for ML-based stealing attacks are closely related to attackers' target information.

% As for the ground truth built up, some of them define the targeted controlled information as class labels, like user input events \cite{spreitzer2018procharvester} and training set members \cite{shokri2017membership}.

\subsection{Feature Engineering} \label{sec_method_feature} %%%%%%%% feature clean/feature extraction %%%%%%%% 
After the datasets are prepared, feature engineering is the subsequent essential phase to generate representative vectors of the data to empower the ML model. The two key points in feature engineering for ML-based attacks consist of dataset cleaning and extracting features.

An obstacle of feature engineering is cleaning the noises and irrelevant information in the raw data. In general, deduplication and interpolation can be used to reduce the noise from accessible resource \cite{diao2016no}. To reduce the noise, a Fast Fourier Transform (FFT) filter and an Inverse FFT (IFFT) filter are applied \cite{liu2015good}. Other popular methods extract refined information and replace redundant information, such as Dynamic Time Warping (DTW) and Levenshtein-distance (LD) algorithms for similarities calculation in time series data \cite{diao2016no,spreitzer2018procharvester, zhang2018level}, Symbolic Aggregate approXimation (SAX) for dimensionality reduction \cite{zhang2018level, patel2002mining}, normalization and discretization for effectiveness \cite{zhang2018level}, and Bag-of-Patterns (BoP) representation and Short Time Fourier Transform (STFT) for feature refinement \cite{zhang2018level,lin2009finding, hojjati2016leave}. 

In order to extract features, it is necessary to analyze and clarify the relationship between the dataset and the targeted controlled information. The relationship determines what kinds of features the attacker should extract. For instance, the inputs and their corresponding confidential values reveal the behaviour of the model stored in cloud service (like Google service). Adversaries choose each query's confidential value as a key feature. Therefore, this relationship can be leveraged to steal an ML model and customer's training samples using reverse-engineering techniques \cite{tramer2016stealing} and the shadow training samples' generation \cite{shokri2017membership}. Using reverse-engineering techniques, the target model's parameters were revealed in \cite{tramer2016stealing}  by finding the threshold where confidential value changes with various inputs. Shadow training samples are intended to be statistically similar to the target training set and are synthesized according to the inputs with high confidence values.

When targeting at user activities, some feature extraction approaches are applied in a kernel dataset for the stealing attack. In \cite{diao2016no, spreitzer2018procharvester}, the diverse foreground apps were characterized by the changes in electrostatic field found in interrupt timing log files stored on Android. The statistics of interrupt timing data are calculated as features \cite{diao2016no}. Feature extraction techniques depend on the type of the useful information. For example, several extraction techniques, including interrupt increment computation, gram segmentation, difference calculation, and the histogram construction, are specialized for the sequential data, like interrupt time series \cite{diao2016no, zhang2018level,xiao2015mitigating}. For the authentication information stealing attack, the ways of defining features are similar to those methods mentioned above \cite{liu2015good, wang2016targeted, melicher2016fast}. One typical method is transforming the characteristics of information as features, such as logical values of the state of sensor \cite{sikder20176thsense}, temporal information accessing memory activities \cite{gras2018translation}, different kinds of PIIs from Internet resources \cite{wang2016targeted, veras2014semantic}, and acceleration segments within a period of time collected from smartwatches' accelerometer \cite{liu2015good}. In addition, manually defining the features based on the attackers' domain knowledge is another popular method \cite{spreitzer2018procharvester, zhang2018level, li2013membership, veras2014semantic}.

\subsection{Attacking the Objective} \label{sec_method_attack} %%%%%%%% two kinds of attacks %%%%%%%%
\begin{figure}[!th]
\centering
\subfloat[The First Attack Mode]{
	\label{subfig:1stMode}
	\includegraphics[width=0.46\textwidth]{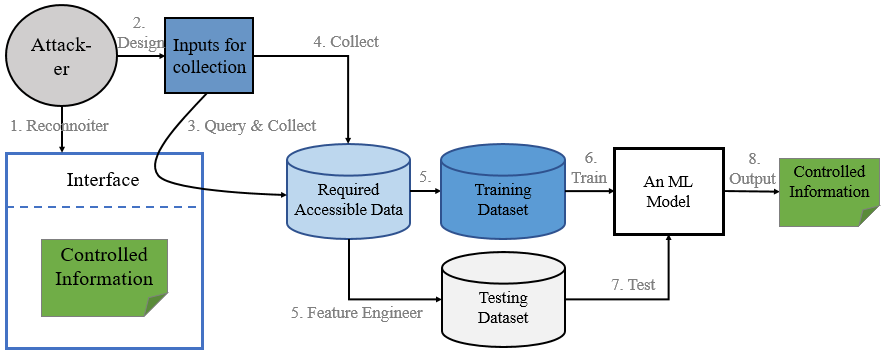}} 
	%&
	\hfill
\subfloat[The Second Attack Mode]{
	\label{subfig:2ndMode}
	\includegraphics[width=0.46\textwidth]{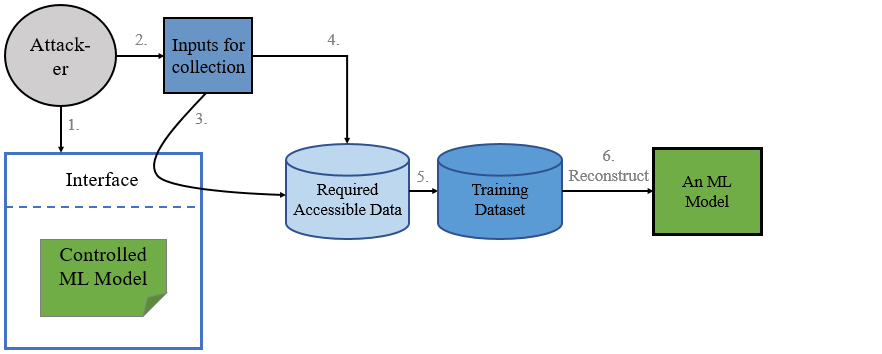}} 
\caption{Two ML-based attack modes: the first mode uses this ML-based model as a weapon to steal controlled information, while this model itself is the target for the second mode. Based on the result from \textit{reconnaissance}, attackers \textit{design} the input queries. \textit{Querying} the target system/service, attackers \textit{collect} required accessible data from the inputs and their query outputs. To set up ground truth, the data are labeled according to the target information. After \textit{feature engineering}, training dataset is built with labels to \textit{train} an supervised ML model. For the first mode, testing samples without labels \textit{test} the model whose \textit{outputs} are the target information. For the second mode, the training dataset is used to \textit{reconstruct} a model which is the attacker's target.} 
\label{fig:attack_mode}
\end{figure}

In this survey, we only consider the ML-based stealing attack as defined in Section~\ref{sec_intro} targeting at user activity information, ML model related information, and authentication information. We summarize the ML-based stealing attack into two attack modes as illustrated in Fig.~\ref{fig:attack_mode}. \skyRe{That is, the initial five actions are identical in both attack modes. These five actions correspond to the first three phases within the MLBSA methodology.} Specifically, the attacker firstly reconnoiters the environment storing targeted controlled information. The environment provides an interface taking users' queries and responding to the queries. The attacker designs the input queries and inquiries the target system/service. As stated in the data collection phase, the inputs and their query results are collected as the required accessible dataset, which reveals the target information. Based on the target information, the ground truth of the dataset is set up in this phase. With proper feature engineering methods, the training dataset is prepared to attack the objective. But the subsequent actions to steal the controlled information using machine learning differ between two attack modes.

For the first attack mode as shown in Fig.~\ref{subfig:1stMode}, the training dataset is used to train an ML model to steal the controlled information. The testing dataset has the same features as the training dataset. Regarding the testing dataset is collected from a victim's system/service, the testing samples are not labeled while querying the attack model. Since the attack model is built to infer the controlled information from these accessible data, the output of the model is the targeted controlled information. This attack mode is applied in the ML-based stealing attack against the user activity information, the authentication information, and training set information. The literature applies ML algorithms to train the classification model such as Logistic Model Tree (LMT) \cite{sikder20176thsense}, $k$-Nearest Neighbors ($k$-NN) \cite{diao2016no, spreitzer2018procharvester, zhang2018level}, Support Vector Machine (SVM) \cite{xiao2015mitigating,zhang2018level}, Naive Bayes (NB) \cite{veras2014semantic,sikder20176thsense}, Random Forest (RF) \cite{liu2015good, sun2016visible, salem2019ml}, Neural Network (NN) \cite{li2013membership,ganju2018property}, Convolutional Neural Network (CNN) \cite{abadi2016deep} and logistic regression \cite{fredrikson2015model,melis2019exploiting}. Apart from these classification models, the probabilistic forecasting model is popular to predict the probability of the real password with a guessing password pattern. There are a few probabilistic algorithms applied, such as Probabilistic Context-Free Grammars (PCFG), Markov model, and Bayesian theory \cite{gras2018translation, veras2014semantic,melicher2016fast}.

For the second attack mode illustrated in Fig.~\ref{subfig:2ndMode}, the training dataset is used to train an ML model while the model itself is the target of the attack. This attack mode is mostly applied in the ML-based stealing attack against the ML model related information. In a black-box setup, stealing the ML model attack aims at calculating the detailed expression of the model's objective function. Reconstructing the original model is essentially a reconstruction attack \cite{fredrikson2015model}. Using the equation-solving and path-finding methods \cite{tramer2016stealing,wang2018stealing}, the inputs and their query outputs for solving the specific objective function expression is interpreted as the training set. Therefore, this attack can be simplify regarded as an ML-based attack. Additionally, based on the attackers' inputs and the query outputs, the training set is synthesized and used to build a substitute model for reconstruction \cite{papernot2017practical,joon2018towards}. Several ML algorithms were applied in the literature, such as decision tree \cite{tramer2016stealing,papernot2017practical}, SVM \cite{tramer2016stealing, wang2018stealing}, NN \cite{tramer2016stealing, wang2018stealing,joon2018towards}, Recurrent Neural Network (RNN) \cite{melicher2016fast}, ridge regression (RR), logistic regression, and linear regression \cite{wang2018stealing, tramer2016stealing,papernot2017practical}. 

Moreover, some popular and publicly available tools can be used to train the ML model for the attack, for example, WEKA and monkeyrunner \cite{diao2016no, sun2016visible}. In summary, despite the model itself is the attacker's objective, the adversary can predict the results which reveal the controlled information in the training data using ML techniques.

\subsection{Evaluation} \label{sec_method_evaluation} %%%%%%%% performance & practical or not %%%%%%%%
% \reviewer{1-4. Section 2.5 has been reviewed many times in the literature, it should be excluded or a real example should be applied for declaring which evaluation criteria are important than others, for example when the FPR is important, the FNR is important ..etc.}

During the evaluation phase, attackers measure how likely they can successfully steal the controlled information. Evaluation metrics differ between two attack modes. As we investigate the ML-based attack under the first attack mode, the attack evaluation measures the performance of the attack model \skyRe{using confusion matrix}. The higher the performance of the model is, the more powerful the weapon the attacker builds. While under the second attack mode, the attack evaluation measures the differences between the attack model and the target model. The attack will be considered more successful when the attack model is more similar to the target model. Evaluation metrics for two attack modes are summarized separately.

For the first attack mode, the attack model is the attacker's weapon. Its performance is measured by effectiveness and efficiency. Specifically, metrics like execution time and battery consumption are used for efficiency evaluation. Most metrics commonly used to measure the effectiveness include accuracy, precision, recall, FPR, FNR, and F-measure, \skyRe{which are derived from the confusion matrix in Table \ref{tab:confusion}}. The evaluation metrics are listed as below.

\begin{table}
\scriptsize
 \caption{Confusion Matrix for Evaluation. As stated in Section~\ref{sec_method_collection}, the outputs of the attack model are the target information. Class $A$ and Class $B$ refer to the subclass of one controlled information, like foreground app $A$ and $B$ in stealing user activity information attack. Additionally, Class $A$ and Class $B$ can also refer to one controlled information and not this information respectively. For example, if Class $A$ is a member of the target training set, then Class $B$ is not member of that set.}
 \label{tab:confusion}
 \begin{tabular}{|c|cc|}
\hline
    \backslashbox{Predicted}{Actual}  & Class $A$ & Class $B$\\
    \hline
    Class $A$ & True Positive (TP) & False Positive (FP) \\
    Class $B$ & False Negative (FN) & True Negative (TN) \\ \hline
\end{tabular}
\end{table}

\begin{itemize}
\item \textbf{Accuracy}: It is also known as \textbf{success rate} and \textbf{inference accuracy} \cite{diao2016no,liu2015good,sun2016visible}. Accuracy means the number of correctly inferred samples to the total number of predicted samples. Accuracy is a generic metric evaluating the attack model's effectiveness. $Accuracy = \frac{TP+TN}{TP+TN +FP+FN}$
\item \textbf{Precision}: It is regarded as one of the standard metrics for attack accuracy \cite{shokri2017membership}. Precision illustrates the percentage of samples correctly predicted as controlled class $A$ among all samples classified as $A$. Precision reveals the correctness of the model's performance on a specific class \cite{spreitzer2018procharvester,sikder20176thsense, li2013membership}, especially when features' values are binary \cite{fredrikson2015model}. \skyRe{$Precision = \frac{TP}{TP +FP}$}
\item \textbf{Recall}: It is regarded as another standard metric for attack accuracy \cite{shokri2017membership}. Recall is also called \textbf{sensitivity} or \textbf{True Positive Rate (TPR)} \cite{sikder20176thsense}. It is the probability of the amount of class A correctly predicted as class $A$. Similar to precision, recall also reveals the model's correctness on a specific class. These two metrics are almost always applied together \cite{spreitzer2018procharvester,sikder20176thsense, fredrikson2015model,shokri2017membership,salem2019ml,ganju2018property,melis2019exploiting}. $Recall = \frac{TP}{TP+FN}$
\item \textbf{F-measure}: This metric or \textbf{F1-score} is the harmonic mean of recall and precision. F-measure provides a comprehensive analysis of precision and recall \cite{sikder20176thsense}. $F-measure = \frac{2\times Recall\times Precision}{Recall+Precision}$
\item \textbf{False positive rate (FPR)}: This metric denotes the proportion of class $B$ samples mistakenly categorized as class $A$ sampled. FPR assesses the model's misclassified samples. $FPR = \frac{FP}{TN +FP}$
\item \textbf{False negative rate (FNR)}: This metric stands for the ratio between class $A$ samples mistakenly categorized as class $B$ samples. Similar to FPR, FNR assesses the model's misclassified samples from another aspect. FPR and FNR are almost always applied together to measure the model's error rate \cite{sikder20176thsense}. $FNR = \frac{FN}{TP+FN}$
\item \textbf{Execution time}: The execution time is used in training the model which indicates the efficiency of the attack model \cite{zhang2018level,diao2016no,zhou2016software}.
\item \textbf{Battery consumption}: It is also known as \textbf{power consumption} \cite{zhang2018level}. Battery consumption refers to the target mobile's battery while the target system is a mobile system \cite{diao2016no,spreitzer2018procharvester,zhang2018level}, which indicates the efficiency of the attack model.
\end{itemize}

For the second attack mode, ML-based attacks of stealing the ML model are assessed with other metrics. This kind of attack is the ML model reconstruction attack. Inherently, the reconstruction attack requires a set of comparison metrics. The target of this kind of attack is an ML model $\hat{f}$ which closely matches the original ML model $f$. Generally, the stolen model $\hat{f}$ will be constructed locally. Its prediction results will be compared to the results of the original model with the same inputs. The applied evaluation metrics are defined and listed below:

\begin{itemize}
\item \textbf{Test error} is the average error based the same test set ($D$) testing at learned model and targeted model \cite{tramer2016stealing}. A low test error means $\hat{f}$ matches $f$ well. $Error_{test}(f,\hat{f}) = \frac{\sum_{x\in D}diff(f(x),\hat{f}(x))}{|D|}$
\item \textbf{Uniform error} is an estimation of the portion of full feature space that the learned model is different from the targeted one, when the testing set ($U$) are selected uniformly \cite{tramer2016stealing}. $Error_{uniform}(f,\hat{f}) = \frac{\sum_{x\in U}diff(f(x),\hat{f}(x))}{|U|}$
\item \textbf{Extraction accuracy} indicates the performance of model extraction attack based on the test error and the uniform error \cite{tramer2016stealing}. $Accuracy_{extraction} = 1 - Error_{test}(f,\hat{f}) = 1 - Error_{uniform}(f,\hat{f})$
\item \textbf{Relative estimation error (EE)} measures the effectiveness of model extraction attack using its learned hyperparameters ($\hat{\lambda}$) contrasting to the original hyperparameters ($\lambda$) \cite{wang2018stealing}. $Error_{EE} = \frac{|\hat{\lambda}-\lambda|}{\lambda}$
\item \textbf{Relative mean square error (MSE)} measures how well the model extraction attack reconstructs the regression models via comparing the mean square error after learning hyperparameters using cross-validation techniques \cite{wang2018stealing}. $Error_{MSE}=\frac{|MSE_{\hat{\lambda}}-MSE_{\lambda}|}{MSE_{\lambda}}$
\item \textbf{Relative accuracy error (AccE)} measures how well the model extraction attack reconstructs the classification models via comparing accuracy error after learning hyperparameters using cross-validation techniques \cite{wang2018stealing}. $Error_{AccE}=\frac{|AccE_{\hat{\lambda}}-AccE_{\lambda}|}{AccE_{\lambda}}$
\end{itemize}

The adversary applies evaluation metrics to determine whether the performance of attack is satisfactory or not. If the value of any metrics does not meet the expectations, adversaries can restart the stealing attack by redefining the targeted controlled information. The stealing attack can be executed incrementally until the attacker gains the satisfactory results.

\section{ML-based Stealing Attacks and Protections} \label{sec_LR}
%%%%% Confirmation Section: Top Conferences Paper for Security %%%%%
%%% Top Conferences Paper %%%
%\begin{table*}[!t] %%----------- Outline of Reviewed Papers ----------%% 
\begin{table*}[!th]
\scriptsize
\caption{Outline of Reviewed Papers (info: information)}
\label{tab:outline_review}
\centering
\begin{tabular}{|c|c|l|l|l|}
%\hline
%    \toprule
\hline
\textbf{Reference} & \textbf{Year} & \textbf{Targeted Info} & \textbf{Accessible Data} & \textbf{Goals} \\
\hline
%    \midrule
\cite{diao2016no} & 2016 & \begin{tabular}[c]{@{}l@{}}Unlock pattern;\\ Foreground app\end{tabular} & \begin{tabular}[c]{@{}l@{}}Hardware interrupt data\end{tabular} & \begin{tabular}[c]{@{}l@{}}Unlock pattern \& foreground app inference attacks via analyzing\\ interrupt time collected from interrupt log file.\end{tabular} \\
\hline
\cite{spreitzer2018procharvester} & 2018 & \begin{tabular}[c]{@{}l@{}}Visited websites;\\ Foreground app\end{tabular} & \begin{tabular}[c]{@{}l@{}}Interrupt data; Network\\  \&Memory process record\end{tabular} & \begin{tabular}[c]{@{}l@{}}Search and attack the kernel records leaking user's specific events\\ (i.e. app starts, website launch, keyboard gesture).\end{tabular} \\
\hline
\cite{zhang2018level} & 2018 & \begin{tabular}[c]{@{}l@{}}Visited websites; \\Foreground app; Map\end{tabular} & \begin{tabular}[c]{@{}l@{}}Memory data; Network\\ source; File system data\end{tabular} & Several side-channel inference attack on iOS mobile device. \\
\hline
\cite{xiao2015mitigating} & 2015 & \begin{tabular}[c]{@{}l@{}}Visited websites;\\ Input keystrokes\end{tabular} & \begin{tabular}[c]{@{}l@{}}Kernel data-structure\\ fields\end{tabular} & \begin{tabular}[c]{@{}l@{}}Protect by injecting noise into the value of kernel data\\ structure values to secure $procfs$.\end{tabular} \\
\hline
\cite{hojjati2016leave} & 2016 & \begin{tabular}[c]{@{}l@{}}Manufacturing\\ activities\end{tabular} & \begin{tabular}[c]{@{}l@{}}Acoustic sensor data;\\Magnetic sensor data\end{tabular} & \begin{tabular}[c]{@{}l@{}}An attack capture acoustic \& magnetic sensor data to steal a \\manufacturing process specification or a design.\end{tabular}\\
\hline
\cite{sikder20176thsense} & 2017 & User activities info & Sensor data & \begin{tabular}[c]{@{}l@{}}Contextual model detect malicious behavior of sensors like leaking.\end{tabular}\\
\hline
\cite{tramer2016stealing} & 2016 & \begin{tabular}[c]{@{}l@{}}Parameters of \\an ML model\end{tabular} & \begin{tabular}[c]{@{}l@{}}Input features \&\\ Query outputs\end{tabular} & \begin{tabular}[c]{@{}l@{}}Model extraction attacks leverage confidence info with\\ predictions against MLaaS APIs in black-box setting.\end{tabular} \\
\hline
\cite{papernot2017practical} & 2017 & \begin{tabular}[c]{@{}l@{}}Internal info of\\ an ML model\end{tabular} & \begin{tabular}[c]{@{}l@{}}Input features \&\\ Query outputs\end{tabular} & \begin{tabular}[c]{@{}l@{}}Build a local model to substitute the target model and use\\ it craft adversarial examples in black-box setting.\end{tabular} \\
\hline
\cite{wang2018stealing} & 2018 & \begin{tabular}[c]{@{}l@{}}Hyperparameters\\ of an ML model\end{tabular} & \begin{tabular}[c]{@{}l@{}}Input features \&\\ Query outputs\end{tabular} & \begin{tabular}[c]{@{}l@{}}Hyperparameters stealing attack via observing minima objective\\ function against MLaaS in black-box setting.\end{tabular}\\
\hline
\cite{joon2018towards} & 2018 & \begin{tabular}[c]{@{}l@{}}Hyperparameters\\ of an ML model\end{tabular} & \begin{tabular}[c]{@{}l@{}}Input features \&\\ Query outputs\end{tabular} & \begin{tabular}[c]{@{}l@{}}Build a metamodel to predict hyperparameters with a given\\ classifier in black-box setting to generate adversarial examples.\end{tabular} \\
\hline
\cite{fredrikson2015model} & 2015 & \begin{tabular}[c]{@{}l@{}}Training data for \\an ML model \end{tabular} & \begin{tabular}[c]{@{}l@{}}Input features \& Query \\outputs \& model structure\end{tabular} & \begin{tabular}[c]{@{}l@{}}Model inversion attacks used confidence info leaking\\ training samples with predictions against MLaaS in two settings.\end{tabular}\\
\hline
\cite{hitaj2017deep} & 2017 & \begin{tabular}[c]{@{}l@{}}Training data for \\an ML model \end{tabular} & \begin{tabular}[c]{@{}l@{}}Input features \& Query \\outputs \& model structure\end{tabular} & \begin{tabular}[c]{@{}l@{}}Online Attack using GAN against collaborative deep learning\\ model leaking user's training sample.\end{tabular}\\
\hline
\cite{shokri2017membership} & 2017 & \begin{tabular}[c]{@{}l@{}}Training data for \\an ML model \end{tabular} & \begin{tabular}[c]{@{}l@{}}Input features \&\\ Query outputs\end{tabular} & \begin{tabular}[c]{@{}l@{}}Membership inference attacks use shadow training technique to\\ leak the specific record's membership of original training set.\end{tabular}\\
\hline
\cite{salem2019ml} & 2019 & \begin{tabular}[c]{@{}l@{}}Training data for \\an ML model \end{tabular} & \begin{tabular}[c]{@{}l@{}}Input features \&\\ Query outputs\end{tabular} & \begin{tabular}[c]{@{}l@{}}Enlarge the scope of membership inference attacks by releasing\\ some key assumptions.\end{tabular}\\
\hline
\cite{ganju2018property} & 2018 & \begin{tabular}[c]{@{}l@{}}The property of \\training set \end{tabular} & \begin{tabular}[c]{@{}l@{}}Input features \& Query \\outputs \& model structure\end{tabular} & \begin{tabular}[c]{@{}l@{}}Infer global properties of the training data unintended to be shared\\ in white-box setting.\end{tabular}\\
\hline
\cite{melis2019exploiting} & 2019 & \begin{tabular}[c]{@{}l@{}}The property of \\training set \end{tabular} & \begin{tabular}[c]{@{}l@{}}Input features \& Query \\outputs \& model structure\end{tabular} & \begin{tabular}[c]{@{}l@{}}Membership inference attacks against collaborative deep learning\\ model leaking others' unintended feature.\end{tabular}\\
\hline
\cite{nasr2018machine} & 2018 & \begin{tabular}[c]{@{}l@{}}Training data for \\an ML model \end{tabular} & \begin{tabular}[c]{@{}l@{}}Input features \&\\ Query outputs\end{tabular} & \begin{tabular}[c]{@{}l@{}}Protect against black-box membership inference attack using\\ an adversarial training algorithm.\end{tabular}\\
\hline
\cite{papernot2017semi} & 2017 & \begin{tabular}[c]{@{}l@{}}Training data for \\an ML model \end{tabular} & \begin{tabular}[c]{@{}l@{}}Input features \&\\ Query outputs\end{tabular} & \begin{tabular}[c]{@{}l@{}}Protect training set of model from leakage with teacher and student\\ models using PATE.\end{tabular}\\
\hline
\cite{lecuyer2017pyramid} & 2017 & \begin{tabular}[c]{@{}l@{}}Training data for \\an ML model \end{tabular} & N/A & Protect training dataset in stored from leakage before training. \\
\hline
\cite{liu2015good} & 2015 & \begin{tabular}[c]{@{}l@{}}Input PINs;\\ User input texts\end{tabular} & \begin{tabular}[c]{@{}l@{}}Acoustic sensor data;\\Accelerometer data\end{tabular} & \begin{tabular}[c]{@{}l@{}}Attack infers users' inputs on keyboards via accelerometer data\\ within user's smartwatch.\end{tabular} \\
\hline
\cite{sun2016visible} & 2016 & \begin{tabular}[c]{@{}l@{}}Input PINs;\\ User input texts\end{tabular} & Audio sensor data & \begin{tabular}[c]{@{}l@{}}Attack infers a user's typed inputs from surreptitious video\\ recordings of a tablet's backside motion.\end{tabular} \\
\hline
\cite{gras2018translation} & 2018 & Cryptographic keys & TLB Cache data & \begin{tabular}[c]{@{}l@{}}TLBleed attack TLBs to leak secret keys about victim's memory\\ activities via reversing engineer and ML strategies.\end{tabular}\\
\hline
\cite{zhou2016software} & 2016 & Secret keys & CPU Cache data & \begin{tabular}[c]{@{}l@{}}Mitigate access-driven side-channel attacks with CacheBar\\ managing memory pages cacheability.\end{tabular}\\
\hline
\cite{wang2016targeted} & 2016 & Password info & \begin{tabular}[c]{@{}l@{}}PII \& leaked password\\ \& site info\end{tabular} & \begin{tabular}[c]{@{}l@{}}Attack with seven mathematical guessing models for seven\\ password guessing scenario using different personal info.\end{tabular}\\
\hline
\cite{veras2014semantic} & 2014 & Password info & \begin{tabular}[c]{@{}l@{}}Corpus \& Site leaked list\end{tabular} & Password guessing attack by analyzing its semantic patterns. \\
\hline
\cite{melicher2016fast} & 2016 & Password info & Corpus library & \begin{tabular}[c]{@{}l@{}}Mitigate against password guessing attack by modeling\\ password guessability in password creation stage.\end{tabular}\\
\hline
\end{tabular}
\end{table*}

This section reviews all the core papers in accordance with the MLBSA methodology presented in the previous section. The review will be undertaken hierarchically according to the structure illustrated in Fig.~\ref{fig:intro_categories}. This section consists of three subsections. The secondary class of Section~\ref{subsec_user} is based on different kinds of accessible data, while Section~\ref{subsec_ML} and Section~\ref{subsec_auth} are grouped by different kinds of controlled targeted information. For each attack category, the attack methods and the corresponding countermeasures are discussed.

To understand the information leakage threat and the stealing attack comprehensively, an outline of relevant high-quality papers from 2014 to 2019 is provided and shown in Table \ref{tab:outline_review}, which lists each paper's targeted controlled information, the accessible data and the goal. In the end, these papers are summarized in Table \ref{tab:sum_review} from four perspectives including the attack, protection, related ML techniques, and evaluation. For each subsection, the key points of the attack are listed, such as ``dataset for an experiment'', ``dataset description'', ``feature engineering (/targeted ML model)'', and ``ML-based attack methods''. Tables \ref{tab:user_kernel} to \ref{tab:authen_psw} summarize all subclasses of the review. In this section, all tables highlight the essential elements of the each ML-based attack. The detailed information of the dataset and source code for these attacks are listed on Github \footnote{\url{https://github.com/skyInGitHub/Machine-Learning-Based-Cyber-Attacks-Targeting-on-Controlled-Information-A-Survey}}.

\subsection{Controlled User Activities Information} \label{subsec_user} %%%%%%%%%%%   User Activities   %%%%%%%%%%% 
\begin{figure*}[!th] %% Attack at user activity information %% 
\centering
\includegraphics[width=.55\textwidth]{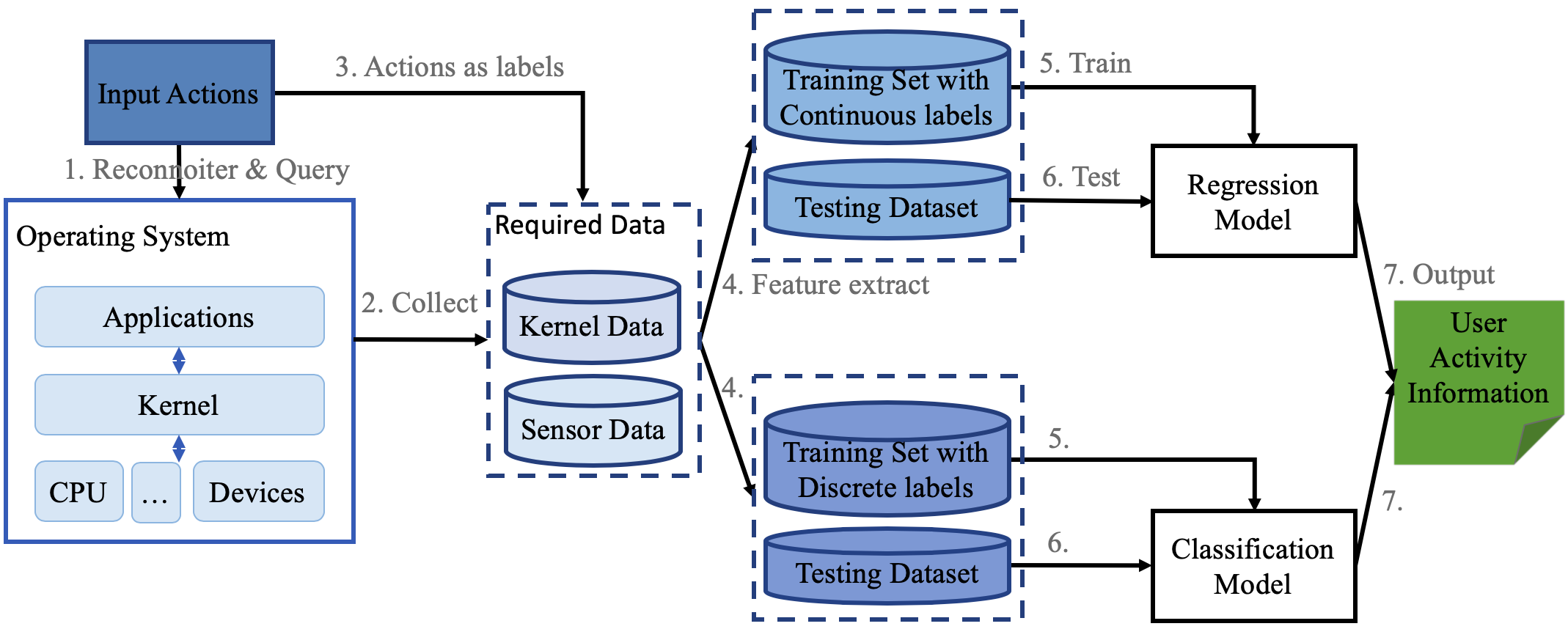}
\caption{The ML-based stealing attack against user activities information. As stated in Section~\ref{sec_method}, \textit{reconnoiter and query} in the reconnaissance phase aim to gain the data that is accessible and valuable to the attack. The required data for this attack can be categorized as kernel data and sensor data. In the data collection phase, these datasets are \textit{collected} and \textit{labeled with input actions} as ground truth. The label's value can be either continuous (i.e.~a series of lines for one unlock pattern \cite{diao2016no}) or discrete (i.e.~various apps \cite{spreitzer2018procharvester}). Upon completion of \textit{extracting features}, the training set will be used to \textit{train} the attack model, and the testing set is prepared to \textit{test} and evaluate the model with its \textit{outputs}. Herein, regression models predict the output as a continuous value (i.e.~swipe lines), whilst classification models predict a discrete value (i.e.~a foreground app).}
\label{fig:LR_activity}
\end{figure*}

It is essential for security specialists to protect user activities information. Not only because the private activities are valuable to adversaries, but also the adversary can exploit some specific activities (i.e.~foreground app) to perform malicious attacks such as the phishing attack \cite{diao2016no}. In general, the attackers pursue two types of data --- kernel data and sensor data, as shown in Fig.~\ref{fig:LR_activity}. As stated in Section~\ref{sec_method}, the works of literature trained their attack models with supervised learning. We organize the reviewed papers according to the MLBSA methodology. The countermeasures against this kind of ML-based stealing attack are discussed at the end of Section~\ref{subsec_user}. According to the utilized kernel data and sensor data, controlled user activities information were stolen through timing analysis and frequency analysis.

\subsubsection{\textbf{Stealing controlled user activities from kernel data}}%%%%% User Activities -> kernel data %%%%% 
The dataset collected from the kernel about system process information is too noisy and coarse-grained to disclose any intelligible and valuable information. Hence, neither specific requirements are required nor strong protections are deployed in accessing these data. However, through analyzing plenty of such data, the adversary could deduce some confidential information about the victim's activities. With the aid of ML algorithms, the effectiveness and efficiency of this attack can be improved significantly. As a result, such an information leakage problem becomes more severe than ever, and the corresponding protection is in great need. The stealing attack utilizes the interrupt sources, such as $procfs$ and the OS-level information, such as memory, network, and file system information.

\begin{table*}[!th] %%%% Stealing Controlled User Activities using Kernel Data %%%% 
\scriptsize
\caption{Stealing Controlled User Activities using Kernel Data}
\label{tab:user_kernel}
\centering
\begin{tabular}{|c|l|l|l|l|}
\hline
\textbf{Reference} & \textbf{Dataset for Experiment} & \textbf{Description} & \textbf{Feature Engineering} & \textbf{ML-based Attack Method} \\
\hline
\cite{diao2016no} & \begin{tabular}[c]{@{}l@{}}Interrupt data for unlock \\pattern and for apps\end{tabular} & \begin{tabular}[c]{@{}l@{}}Collect from $procfs$\end{tabular} & \begin{tabular}[c]{@{}l@{}}Deduplication; Interpolation; \\Interrupt Increment Computation; \\Gram Segmentation; DTW \end{tabular} & \begin{tabular}[c]{@{}l@{}}HMM with Viterbi \\algorithm; $k$-NN classifier\\ with DTW\end{tabular}\\
\hline
\cite{spreitzer2018procharvester} & \begin{tabular}[c]{@{}l@{}}Time series for apps,\\ website,keyboard guests\end{tabular} & \begin{tabular}[c]{@{}l@{}}Collect from $procfs$\end{tabular} &\begin{tabular}[c]{@{}l@{}} Automatically extract with \\ $tsfresh$; DTW\end{tabular}  & \begin{tabular}[c]{@{}l@{}}Viterbi algorithm with DTW;\\ SVM classifier with DTW\end{tabular}\\
\hline
\cite{zhang2018level} & \begin{tabular}[c]{@{}l@{}}1200 x 6 time series of\\ data about app;\\ 1000 website traces\end{tabular} & \begin{tabular}[c]{@{}l@{}}120 apps(App Store+iOS )\\+10 trace x 6 time series;\\ 10 traces for each website\end{tabular} & \begin{tabular}[c]{@{}l@{}}Manually defined;\\ SAX, BoP representation\end{tabular} & \begin{tabular}[c]{@{}l@{}}SVM classifier;\\ $k$-NN classifier\\ with DTW\end{tabular}\\
\hline
\cite{xiao2015mitigating} & \begin{tabular}[c]{@{}l@{}}Consecutively reading \\data; Resident size field data\end{tabular} & \begin{tabular}[c]{@{}l@{}}Collect from $procfs$\end{tabular} & \begin{tabular}[c]{@{}l@{}}N/A; Construct a histogram binned\\ into seven equal-weight bins\end{tabular} & SVM classifiers\\
\hline
\end{tabular}
\end{table*}

%% No Pardon for the Interruption: New Inference Attacks on Android Through Interrupt Timing Analysis \cite{diao2016no}
%% ProcHarvester: Fully Automated Analysis of Procfs Side-Channel Leaks on Android \cite{spreitzer2018procharvester}
\textit{Stealing User Activities with Timing Analysis:} The security implications of the kernel information were evaluated in \cite{diao2016no, spreitzer2018procharvester} through integrating some specific hardware components into Android smartphones. During the reconnaissance phase, user activities information records the user's interactions with hardware devices, before responded by the kernel layer. The targeted user activities in \cite{diao2016no} were unlock patterns and foreground apps. Moreover, users' browsing behavior was targeted by the attacker in \cite{spreitzer2018procharvester}. One kind of kernel data was accessible to legitimate users, which logged the time series of the hardware interrupt information can reveal previous activities. Specifically, the reported interrupts imply the real-time running status of some specific hardware (i.e.~touchscreen controller). However, accessing the similar process-specific information had been continuously restricted since Android 6, and the interrupt statistics became unavailable in Android 8 \cite{spreitzer2018procharvester}. Different Android versions contain different kinds of process information which is accessible to legitimate users without permissions under $proc$ filesystem ($procfs$). Thus, an app was developed in \cite{spreitzer2018procharvester} to search all accessible process information under $procfs$. The time series of these accessible data could distinguish the event of interests including unlocked screen, the foreground app, and the visited website. Reconnaissance showed the value of time series of data in $procfs$.

During the data collection, the interrupt time logs were collected by pressing and releasing the touchscreens in \cite{diao2016no}. Specifically, for the versions prior to Android 8, a variety of interrupt time series recording the changes of electrostatic field from touchscreen were gathered as one dataset for stealing the user's unlock pattern. Another dataset was built for stealing foreground apps' information by recording the time series of starting the app from accessible sources like interrupts from the Display Sub-System \cite{diao2016no} and the virtual memory statistics \cite{spreitzer2018procharvester}. Moreover, the time series of some network process information fingerprinted online users. These fingerprints were gathered as the dataset for stealing user's web browsing information. Different sets of time series were prepared with respect to the information of different user activities.

In terms of feature engineering, the attacker can analyze the process information on $procfs$ to study the characteristics of the user's unlock pattern, foreground app status, and the user's browsing activity. The datasets were firstly processed by deduplication, interpolation and increment computation. The distinct features of three datasets were constructed via several methods such as segmentation, similarity calculation and DTW. An automatic method named $tsfresh$ \cite{christ2016distributed} was utilized for feature extraction. Subsequently, for the stealing attack targeting unlock patterns, a Hidden Markov Model (HMM) was used to model the attack to infer the unlock patterns through the Viterbi algorithm \cite{forney1973viterbi}. The evaluation results showed that its success rate outperformed the random guessing significantly. Targeting at foreground apps, the processed data was used to train a $k$-NN classifier. For the evaluation, the results showed that the classifier had high accuracy, which achieved 87\% on average in \cite{diao2016no} and 96\% in \cite{spreitzer2018procharvester}. To reveal the user's browsing activities, SVM classifier was used to mount the attack. The results showed that both precision and recall values were above 80\% in \cite{spreitzer2018procharvester}. Among these three attack scenarios, the consumption of battery and time were acceptable (less than 1\% and shorter than 6 minutes). The ML-based stealing attack showed its effectiveness with less consumption in time and battery.

%New attacks could be brought out from another interrupt source, although the similar interrupts information was not public to normal users in Mac OS X/iOS environment \cite{diao2016no}. Although the interrupts information was public on Microsoft Windows platforms, the attack methods might be different because of different implementations. On the other hand, since the detection system proposed by \cite{zhang2015leave} could not protect a user from these two attacks, another two defense mechanisms were introduced, namely fine-grained access control on kernel files and decreasing the resolution of interrupt data. Furthermore, a few phone models were tested for the attacks such as Google Nexus 6, Sony Xperia Z3, Samsung Galaxy S and so on. In \cite{spreitzer2018procharvester}, ProcHarvest was introduced which could automatically detect known information leaks by analyzing the $proc$ files. Attackers need to search other accessible source to evade the current protection.

%% OS-level Side Channels without Procfs: Exploring Cross-App Information Leakage on iOS \cite{zhang2018level}
\textit{Stealing User Activities with iOS Side-channel Attack:} The OS-level side-channel attacks were investigated in \cite{zhang2018level} on iOS-based mobile devices (Apple), which stole user activities information. In iOS systems, one popular side-channel attack vector of Linux system about the process information --- $procfs$ --- is inaccessible, which hinders the aforementioned attacks from leaking the sensitive information. Attackers have actively looked for new resources to exploit in the Operating System level.

In the reconnaissance phase, several attack vectors, which are feasible to Apple, were applied to perform cross-app information leakage \cite{zhang2018level}. Specifically, three iOS vectors enabled apps accessing the global usage statics without requiring any special permissions to bypass the timing channel: the memory information, the network-related information, and the file system information. The attacker aimed to steal the user activities' information (such as foreground apps, visited websites and map searches) and in-app activities (such as online transactions). To collect data for an ML-based attack, attackers manually collected several data traces for interesting events like foreground apps, website footprints and map searches. To improve the performance of such an inference attack, the information collected from multiple attack vectors was combined and fed into the ML models. Particularly, time series data from the targeted vectors were exploited frequently. As for feature engineering and the stealing attacks, ML frameworks were utilized to exfiltrate the user's information from accessible vectors \cite{zhang2018level}. The changes of the time series reflected in the difference between two consecutive data traces. The feature processing methods were applied to transform the sequences into the Symbolic Aggregate approXimation (SAX) strings \cite{patel2002mining} and to construct the Bag-of-Patterns (BoP) of the sequences. In \cite{zhang2018level}, two ML-based attacks with a large amount of data were presented --- classifying the user activities and detecting the sensitive in-app activities. An SVM classifier was trained and tested for the former attack. The Viterbi algorithm \cite{forney1973viterbi} with DTW was utilized for the latter attack. In terms of the evaluation of the first attack stealing three users' activities, the foreground app classification accuracy achieved 85.5\%, Safari website classification accuracy reached 84.5\%, and the accuracy accomplished 79\% in inferring map searches. The proposed attacks could be trained on the attacker's device and tested on other devices such as the victim's devices. Meanwhile, the power consumption was acceptable with only 5\% extra power used in an hour, while the attacks' execution time was tolerable as well (within 19 minutes). In the context of stealing user activities information, ML-based attacks exploited the OS-level data with time series analysis.

%Finally, the researchers of \cite{zhang2018level} offered the mitigation suggestions including removing relative APIs, limiting the API query rate, reducing return value granularity, run-time detection, privacy-preserving statistics reporting to kernel like \cite{xiao2015mitigating}, and eliminating the timing channel towards the file system. Apple has deployed some of the suggestions into iOS 11, MacOS High Sierra 10.13 and later versions. In a word, \cite{zhang2018level} revealed several attack vectors within iOS leading the cross-app information leakage. These attacks can be mitigated by disrupting accessible data or preventing them from access. 

%% Mitigating Storage Side Channels Using Statistical Privacy Mechanisms \cite{xiao2015mitigating}
\textit{Protection using Privacy Mechanism:} An attack exploiting the kernel process information \cite{diao2016no} via decreasing the data's resolution was defended by \cite{xiao2015mitigating}. A differential privacy (DP) mechanism was utilized to prevent the attackers from gaining any useful storage information. Additionally, the noise was injected into the kernel data-structure values in order to protect the contents within the $procfs$ with quantifiable security. Specifically, a generalized differential-privacy was applied to quantify the distance between two series of $procfs$ information, which is essential to infer the information about user activities. To retain the utility of $procfs$, the invariants of these noised fields were reestablished on the noised output, in order to assist the applications that depend on them. This method is named as $dpprocfs$, a modified $procfs$, preventing the attacker from utilizing the value reported via $procfs$ interfaces. As a result, $dpprocfs$ showed a reliable security guarantee resistant to the stealing attack with noise injection and generalization.

Two particular attacks were proposed to evaluate the protection approach in \cite{xiao2015mitigating}, including defending against the keystroke timing attack and mitigating the website inference. The former attack collected the data from the kernel layer to obtain the status of keystroke actions. When attacking the objective, an SVM classifier was trained to recognize what keystrokes occurred \cite{xiao2015mitigating}. The latter attack collected the data from the browser' memory footprints in $procfs$ regarding the top-10 websites. By constructing a histogram of the counts of visited websites, records with bin counts are extracted as features and the websites as labels. This dataset was used to train and test an SVM classifier. As for the evaluation of applying the privacy-preserved kernel records in $procfs$, both of its security and utility were assessed. From the security aspect, by infusing noise to the kernel data-structure values, the success rate of the keystroke timing attack was reduced significantly (around 44\%). Similarly, in the website inference scenario, the accuracy of this attack declined significantly. From the utility aspect, several $procfs$ queries were used to test the top $k$ processes in $dpprocfs$ which were influenced by noised kernel data-structure values. The relative errors were the modest (within 5\%) for both scenarios at a low level, while the rank accuracy on top $k$ processes for monitoring and diagnosis tests were at a high level (all over 80\% for top-10). However, the utility would decay if the number of queries grew and a large number of noises were added. The effectiveness of noise injection method was unclear against attacks exploiting other stored information. In summary, a defense proposed by \cite{xiao2015mitigating} added noise to accessible data against the information inference attacks leveraging the kernel data breach. 

\subsubsection{\textbf{Stealing controlled user activities using sensor data}}%%%%% User Activities -> sensor data %%%%%
The stealing attack using sensor data should be studied seriously by the defenders, not only from the application of effective ML mechanisms, but also from the popularity of sensing enabled applications. Currently, there are an increasing number of apps using various sensor information on smart devices to improve the efficiency and user experience \cite{lane2010survey,lane2011enabling, park2011effect,yu2010exploration, macias2013mobile}. The sensor information can reveal the controlled information indirectly as demonstrated in this stealing attack, such as acoustic and magnetic data. 
\begin{table*}[!th] %%%% Stealing Controlled User Activities using Sensor Data %%%% 
\scriptsize
\caption{Stealing Controlled User Activities using Sensor Data}
\label{tab:user_sensor}
\centering
\begin{tabular}{|c|l|l|l|l|}
\hline
\textbf{Reference} & \textbf{Dataset for Experiment} & \textbf{Description} & \textbf{Feature Engineering} & \textbf{ML-based Attack Method} \\
\hline
\cite{hojjati2016leave} & Audio signature dataset & \begin{tabular}[c]{@{}l@{}}Recorded with a phone put\\ within 4 inches of the printer\end{tabular} & \begin{tabular}[c]{@{}l@{}}STFT,\\ noise normalization\end{tabular} & A regression model\\
\hline
\cite{sikder20176thsense} & Sensor dataset & \begin{tabular}[c]{@{}l@{}}Sensor data collected benign\\ and malicious activities\end{tabular} & N/A & \begin{tabular}[c]{@{}l@{}}Markov Chain, NB, LMT, \\ (alternative algorithms e.g.~PART)\end{tabular}\\
\hline
\end{tabular}
\end{table*}

%% Leave Your Phone at the Door: Side Channels that Reveal Factory Floor Secrets \cite{hojjati2016leave}
\textit{Stealing Machine's Activities with Sensor-based Attack:} A side-channel attack was proposed by \cite{hojjati2016leave} on manufacturing equipment exploiting sensor data collected by mobile phones, which revealed its design and the manufacturing process. That is, the attacker managed to reconstruct the targeted equipment. As a result of reconnaissance, the security threat of the manufacturing sector was indicated. In detail, the adversary placed an attack-enabled phone near the targeted equipment like a 3D printer. The accessible acoustic and magnetic information reflected the product's manufacturing activities indirectly. During data collection phase, the acoustic and magnetic sensors embedded in the phone would record audio and gain magnetometer data from the manufacturing equipment. The magnetometer data was transferred into a type of acoustic information. Then these acoustic signal information was combined as the training dataset. Hence, acoustic and magnetic data can be leveraged by the attack.

After the dataset was gathered, the ML-based attack in \cite{hojjati2016leave} was completed by feature engineering, attacking with model training, and evaluation. The features were extracted from the audio signal's frequency with the help of STFT and the noise normalization \cite{hojjati2016leave}. With features constructed, the product's manufacturing process could be inferred by an ML model, especially for 3D printers. A regression algorithm was used to train the ML model for this attack. In the experiments, the adversaries tested the effectiveness of the reconstruction of a star, a gun and an airplane by a 3D printer. All products were reconstructed except the airplane, which was more like a ``fish mouth'' \cite{hojjati2016leave}. As for the difference of angles between original product and the reconstructed one, the differences of all angles were within one degree in average, which was acceptable. A defense of this kind of attack was proposed by \cite{hojjati2016leave}. The protection method obfuscated the acoustic leakage by adding the noise (i.e.~play recordings) during production. Sensor-based attacks built up model by analyzing the frequency of the manufacturing equipment, but noise injection can mitigate this attack to some extent.

%% 6thSense: A Context-aware Sensor-based Attack Detector for Smart Devices \cite{sikder20176thsense}
\textit{Context-aware Sensor-based Detector:} A framework named 6thSense in \cite{sikder20176thsense} detects the malicious behavior of sensors leaking the information about user activities. Three sensor-based threats were highlighted as --- using sensors to send a message to activate a malicious app, using sensors to leak information to any third party, and using the sensor to steal data to deduce a particular device mode (like sleeping). The last two threats related to the attack intended to steal user activities information via sensor data. The ML-based method was used to prevent these threats by detecting the abnormal recordings of sensor information. The dataset was collected from nine sensors while 50 users conducted nine activities with malicious and benign apps. The training samples were collected to present not only the user activities leakage behaviors but also the usual changes within the sensors. Three ML algorithms listed in Table \ref{tab:user_sensor} were adopted to create contextual models in the 6thSense framework to differentiate the benign or malicious sensor behaviors. According to an empirical evaluation, 6thSense achieved high classification performance with over 96\% accuracy. ML models were trained to detect any sensor's malicious behavior caused by sensor-based attacks.

\subsubsection{\textbf{Summary}}%%%%%------------ Summary ------------%%%%% 
ML-based attacks in Section~\ref{subsec_user} steal user activities information from operating systems. According to the data sources, there are two kinds of attacks --- using kernel data and using sensor data. Kernel data reveals some system-level behaviors of the target system, while sensor data reflects the system's reactions on its specific functionality used by users \cite{diao2016no}. The kernel data is analyzed by the adversary from a time dimension, while the sensor data is exploited with frequency analysis.

\textit{Countermeasures:} Regarding the protection mechanism, differential privacy is an important method for the attacks stealing user activities information. In \cite{xiao2015mitigating, zhang2015leave}, an example applied the noise to an accessible data source (like Android kernel log files). Another kind of solution is to restrict access to accessible data \cite{zhang2018level}. It is also effective to build a model to detect potential stealing threats like in \cite{sikder20176thsense}. The in-depth research in protecting against user activities information can explore the differential privacy appliance or a management system design for kernel files and sensor data. Noise injection and access restriction are two effective protections, and the detection can alert the stealing attack.

\subsection{Controlled ML Model Related Information} \label{subsec_ML} %%%%%%%%%%%%   ML Model   %%%%%%%%%%%% 
\begin{figure*}[!th] %% Attack at ML Model Related information %% 
\centering
\includegraphics[width=.55\textwidth]{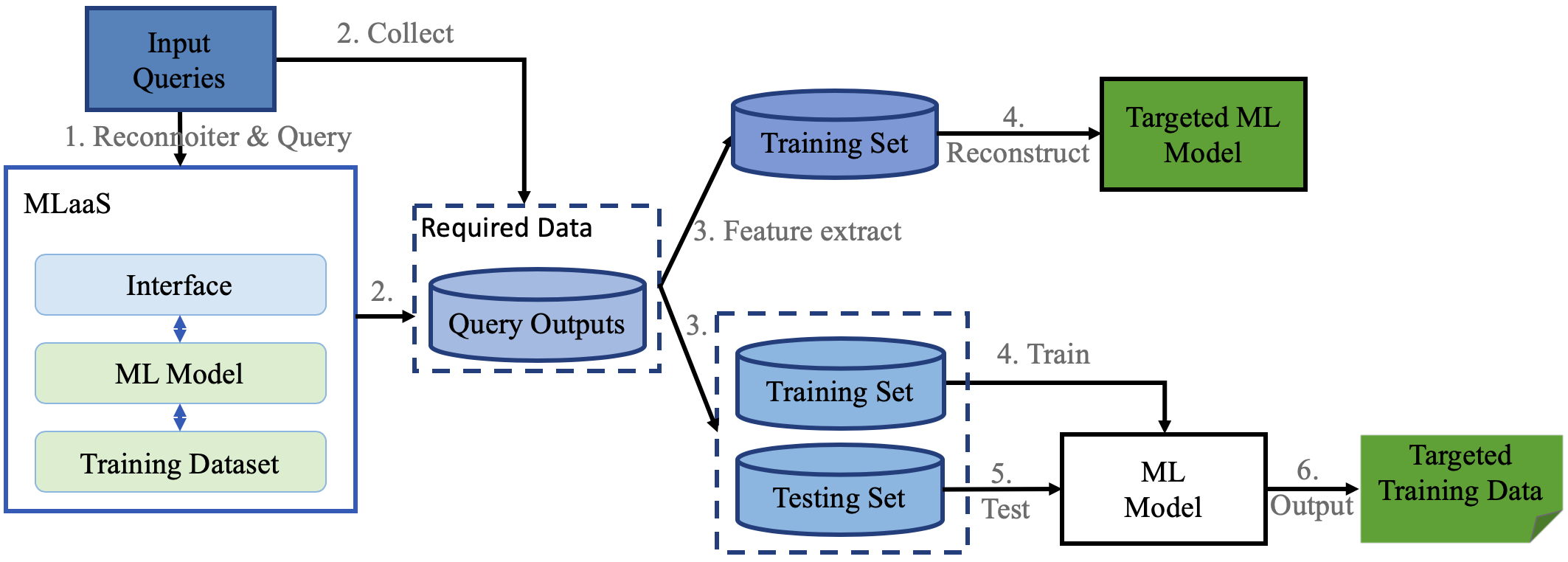}
\caption{The ML-based stealing attack against ML model related information. In this category, ML-based attacks aim at stealing the training samples or the ML model. Stealing the controlled training sample attacks use an ML model to determine whether the input sample is contained in the target training set.}
\label{fig:LR_ml}
\end{figure*}

ML model related information consists of the model description, training data information, testing data information, and testing results. In this subsection, the ML model and users' uploaded training data are the targets, which are stored in the cloud. By querying the model via MLaaS APIs, the prediction/classification results are displayed. The model description and training data information are controlled, otherwise, it is easy for an attacker to interpret the victim's query result. As most of ML services charge users per query \cite{AmazonML,MicrosoftML, GoogleML}, this kind of attack may cause huge financial losses. Additionally, several ML models including neural networks are suffered from adversarial examples. Adding small but intentionally worst-case perturbations to inputs, adversarial examples result in the model predicting incorrect answers \cite{goodfellow2015explaining}. By revealing the knowledge of either the model's internal information or its training data, the stealing attack can facilitate the generation of adversarial examples \cite{papernot2017practical,joon2018towards}.
The generalized attack in this category is illustrated in Fig.~\ref{fig:LR_ml}. Leveraging the query inputs and outputs, the model description can be stolen by using a model extraction attack or a hyperparameter stealing attack, and the training samples can be stolen by using the model inversion attack, GAN attack, membership inference attack, and property inference attack. %The countermeasures mitigating these attacks are summarized at the end of this subsection.

\subsubsection{\textbf{Stealing controlled ML model description}}%%%%%----- ML Model -> model's parameter ----%%%%% 
It is important to protect the confidentiality of ML models online. If the ML model's knowledge description was stolen, the profit of the MLaaS platform may diminish because of its pay-per-query deployment \cite{tramer2016stealing}. If spam or fraud detection are based on ML models \cite{biggio2013evasion, huang2011adversarial, laskov2014practical, lowd2005adversarial}, understanding the model means that adversaries can evade detection \cite{lowd2005adversarial}. A specific ML model is defined by two important elements including ML algorithm's parameters and hyperparameters. Parameters are learned from the training data by minimizing the corresponding loss function. Additionally, hyperparameters aid to find the balance within objective function between its loss function and its regularization terms, which cannot be learned directly from the estimators. Since the model is controlled, its parameters and hyperparameters should be deemed confidential by nature. Stealing these model descriptions, the main approaches are equation-solving, patch-finding and linear least square methods.
\begin{table*}[!th] %% Stealing Controlled ML Model Description %% 
\scriptsize
\caption{Stealing Controlled ML Model Description}
\label{tab:ML_description}
\centering
\begin{tabular}{|c|l|l|l|l|}
\hline
\textbf{Reference} & \textbf{Dataset for Evaluation} & \textbf{Description} & \textbf{Targeted ML Model} & \textbf{Attack Methods} \\
\hline
\cite{tramer2016stealing} & \begin{tabular}[c]{@{}l@{}}Circles, Moons, Blobs, \\ 5-Class \cite{tramer2016stealing}; \\ Steak Survey \cite{UCI}, \\ GSS Survey \cite{smith2012general}, \\ Adult (Income/race) \cite{UCI}, \\ Iris \cite{UCI}, \\ Digits \cite{pedregosa2011scikit}, \\ Breast Cancer \cite{UCI}, \\ Mushrooms \cite{UCI}, \\ Diabetes \cite{UCI}\end{tabular} & \begin{tabular}[c]{@{}l@{}}Synthetic, 5,000 with 2 features,\\ Synthetic,1000 with 20 features,\\ 331 records with 40 features,\\ 16,127 records with 101 features,\\ 48,842 records with 108/105 features,\\ 150 records with 4 features,\\ 1,797 records with 64 features,\\ 683 records with 10 features,\\ 8,124 records with 112 features,\\ 768 records with 8 features\end{tabular} & \begin{tabular}[c]{@{}l@{}}Logistic Regression;\\ Decision Tree;\\ SVM;\\ Three-layer NN\end{tabular} & \begin{tabular}[c]{@{}l@{}}Equation-solving \\attack; Path-finding\\ attack\end{tabular}\\
\hline
\cite{papernot2017practical} & \begin{tabular}[c]{@{}l@{}}MNIST \cite{lecun1998mnist}, \\ GTSRB \cite{stallkamp2012man} \end{tabular} & \begin{tabular}[c]{@{}l@{}}70,000 handwritten digit images,\\ 49,000 traffic signs images\end{tabular} & \begin{tabular}[c]{@{}l@{}} DNN; SVM; $k$-NN;\\ Decision Tree;\\ Logistic Regression\end{tabular} & \begin{tabular}[c]{@{}l@{}}Jacobian-based Dataset\\ Augmentation \end{tabular} \\
\hline
\cite{wang2018stealing} & \begin{tabular}[c]{@{}l@{}}Diabetes \cite{UCI},\\ GeoOrig \cite{UCI}, \\ UJIIndoor \cite{UCI};\\ Iris \cite{UCI},\\ Madelon \cite{UCI}, \\ Bank \cite{UCI}\end{tabular} & \begin{tabular}[c]{@{}l@{}}442 records with 10 features,\\ 1,059 records with 68 features,\\ 19,937 records with 529 features;\\ 100 records with 4 features;\\ 4,400 records with 500 features;\\ 45,210 records with 16 features\end{tabular} & \begin{tabular}[c]{@{}l@{}}Regression algorithms;\\ Logistic regression \\algorithms; SVM; NN\end{tabular} & Equation solving\\
\hline
\cite{joon2018towards} & MNIST \cite{lecun1998mnist} & 70,000 handwritten digit images & NNs & Metamodel methods \\
\hline
\end{tabular}
\end{table*}

%% Stealing Machine Learning Models via Prediction APIs \cite{tramer2016stealing}
\textit{Stealing Parameters Attack:} Model extraction attacks targeting ML models of the MLaaS systems were described in \cite{tramer2016stealing}. The goal of the model extraction attacks was constructing the adversary's own ML model which closely mimics the original model on the MLaaS platform. That is, the constructed ML model can duplicate the functionality of the original one. During the reconnaissance, MLaaS allows clients to access the predictive model in the black-box setting through API calls. That is, the adversary can only obtain the query results. Most MLaaS provides information-rich query results consisting of high-precision confidence values and the predicted class labels. Adversaries can exploit this information to perform the model extraction attack. The first step was collecting confidence values with query inputs. Feature extraction needs to map the query inputs into a feature space of the original training set. Feature extraction methods were applied for categorical and numerical features as listed in Table \ref{tab:ML_description}. Equation-solving and patch-finding attacks were used to calculate the objective function of the targeted model. Three popular ML models were targeted listed in Table \ref{tab:ML_description}, while two online services namely BigML \cite{BigML} and Amazon ML \cite{AmazonML} were compromised as case studies. The key processes of model extraction attacks include query input design, confidence values collection, and attack with equation-solving and patch-finding.

Stealing the model's parameters, the equation-solving attack and patch-finding attack are illustrated in detail. Regarding the attack mentioned in \cite{tramer2016stealing}, the equation-solving attacks can extract confidence values from all logistic models including logistic regression and NNs, whereas the patch-finding attacks work on decision trees model. The equation solving was based on the large class probabilities for unknown parameters and then calculated the model. Specifically, the objective function of the targeted ML model was the equation which adversaries aimed to solve. With several input queries and their predicted probabilities, the parameters of the objective function were calculated. The patch-finding attack exploited the ``ML API particularities'' to query specific inputs to traversal the decision trees \cite{tramer2016stealing}. A path-finding algorithm and a top-down approach helped locate the target model algorithm to reveal paths of the tree. In this way, the detailed structure of the targeted decision tree classifier was reconstructed. The performance of such an attack was evaluated based on the extraction accuracy. The online model extraction attack targeted the decision tree model which was set up by the users on BigML \cite{BigML}. The accuracy was over 86\% irrespective of the completeness of queries. In another case study targeting the ML model on Amazon services, the attacker reconstructed a logistic regression classification model. The results showed that the cost of this attack was acceptable in terms of time consumption (less than 149s) and the price charged (\$0.0001 per prediction). The model was successfully learned by calculating the parameters.

%Furthermore, the model of MLaaS is still vulnerable to model extraction attack with an adaptive algorithm \cite{lowd2005adversarial}, even if its output does not contain any confidence values (i.e. only remains class labels) \cite{tramer2016stealing}. Some protections mentioned in \cite{tramer2016stealing} are helpful to some extent, namely, predictive API minimization like forbidden incomplete queries and rounding confidences \cite{fredrikson2015model}, differential privacy \cite{dwork2008differential,li2013membership}, and ensemble methods returning an aggregated number of individual models' predictions. The model with the protection cannot be reconstructed of high accuracy.

% Practical Black-Box Attacks against Machine Learning \cite{papernot2017practical}

Apart from reconstructing the exact model parameters, another model extraction attack reveals the model's internal information by building a substitute model in \cite{papernot2017practical}. Herein, the substitute model shares similar decision boundaries with the target model. During the reconnaissance, adversaries can only obtain labels predicted by the target model with given inputs. To train this substitute model, a substitute dataset is collected using a synthetic data generation technique named Jacobian-based Dataset Augmentation with a small initial set \cite{papernot2017practical}. Specifically, the ground truth of a synthetic data has the label predicted by the target model, but the architecture is selected based on the understanding of the classification task. The best synthetic training set is determined by the substitute model's accuracy and the similarity of decision boundaries. To approximate the target model's boundaries, the Jacobian matrix is used to identify the directions of the changes in the target model's output. Hence, the model can be reconstructed as a substitute model.

%% Stealing Hyperparameters in Machine Learning \cite{wang2018stealing}
\sky{\textit{Stealing Hyperparameters Attack:} Stealing hyperparameters in the objective function of the targeted MLaaS model may result in gaining financial benefits \cite{wang2018stealing}.} The investigated MLaaS models can be regarded as black-box providing query results only, like Amazon ML \cite{AmazonML} and Microsoft Azure Machine Learning \cite{MicrosoftML}. By analyzing the model's training process, the attacker successfully learned the model's parameters when the objective function reached its minimum value. That is, the gradient of objection function at the model parameters should be the vector whose entries are all close to zeros. Hence, the hyperparameters were learned covertly with a system of linear equations, when the gradient was set to vector of zero.

A threat model was proposed by \cite{wang2018stealing} where the attacker acted as a legitimate user of MLaaS platform. Some popular ML algorithms used by the platform were analyzed in Table \ref{tab:ML_description}. That is, the attacker knew the ML algorithm in advance. Given the learned model's parameters, the attacker set the gradient vector of the objective function of the non-kernel/kernel algorithm. By solving this equation with the linear least square method, the attacker found the hyperparameters. In some black-boxed MLaaS models, the attacker applied the model parameter stealing attacks relied on equation-solving in \cite{tramer2016stealing} to learn the parameters as an initial step. Thus, even though the parameters were hidden, the attacker could steal the hyperparameters. Therefore, the target model was reconstructed successfully.

To evaluate the effectiveness of this stealing hyparameters attack \cite{wang2018stealing}, several real-world datasets listed in Table~\ref{tab:ML_description} were used. Additionally, a set of hyperparameters whose span covered a large range were predefined. Apart from these, a scikit-learn package was applied to implement different ML models and worked out the values of each hyperparameter. For empirical evaluation, relative mean square error (MSE), relative accuracy error (AccE), and relative estimation error (EE) were applied. The results showed a high accuracy of attack performance with all estimation errors were below 10\%. The good performance indicated that the attacker successfully stole the target model.

Regarding a target model as a black-box, its hyperparameters can also be learned by building another metamodel which takes various classifiers' input and output pairs as training data \cite{joon2018towards}. Firstly, by observing outputs of the target model with given inputs, a diverse set of white-box models need to be trained by varying values of various hyperparameters (i.e.~activation function, the existence of dropout or maxpooling layers, and etc). More importantly, these white-box models are expected to be similar to the target model. The training set of the metamodel can be collected by querying inputs over these white-box models, while the ground truth label should be the hyperparameter's value used by the corresponding white-box model. Afterwards, by querying the target model, its hyperparameter can be predicted given its output to the metamodel. Stealing hyperparamters attack is complementary to the stealing parameters attack.

\subsubsection{\textbf{Stealing controlled ML model's training data}}%%%%%--- ML Model -> model's training data ---%%%%% 
Another type of controlled information about MLaaS product is the training data. Training data is not only useful to construct the model using ML algorithms provided by an MLaaS platform, but also sensitive as the records can be private information \cite{fredrikson2015model, fredrikson2014privacy}. For example, a user's health diagnostic model is trained by personal healthcare data \cite{shokri2017membership}. Hence, the confidentiality of the model's training data should be protected. The ML-based stealing attacks include the model inversion attack, GAN attack, membership inference attack, and property inference attack. Moreover, two protections are demonstrated: One uses adversarial regularization against membership inference attack, while another utilizes count featurization for protecting the models' training data.

\begin{table*}[!th] %%%% Stealing Controlled ML Model's Training Data %%%%
\scriptsize
\caption{Stealing Controlled ML Model's Training Data. \sky{A method was proposed in \cite{nasr2018machine} to prevent training set leakage against the membership inference attack \cite{salem2019ml} which provides a simple attack without using shadow models.} Because the methods in \cite{papernot2017semi,lecuyer2017pyramid} were proposed for only protecting training data, the feature engineering and methods for the ML-based attack are omitted.}
\label{tab:ML_training}
\centering
\begin{tabular}{|c|l|l|l|l|}
\hline
\textbf{Reference} & \textbf{Dataset for Experiment} & \textbf{Description} & \textbf{Feature Engineering} & \textbf{ML-based Attack Method} \\
\hline
\cite{fredrikson2015model} & \begin{tabular}[c]{@{}l@{}}FiveThirtyEight survey,\\ GSS marital happiness survey\end{tabular} & \begin{tabular}[c]{@{}l@{}}553 records with 332 features,\\ 16,127 records with 101 features\end{tabular} & N/A & \begin{tabular}[c]{@{}l@{}}Decision Tree,\\ Regression model\end{tabular}\\
\hline
\cite{hitaj2017deep} & \begin{tabular}[c]{@{}l@{}}MNIST \cite{lecun1998mnist},\\ AT\&T \cite{samaria1994parameterisation} \end{tabular} & \begin{tabular}[c]{@{}l@{}}70,000 handwritten digit images,\\ 400 personal face images\end{tabular} & \begin{tabular}[c]{@{}l@{}}Features learned\\  with DNN\end{tabular} & \begin{tabular}[c]{@{}l@{}}Convolutional Neural \\ Network (CNN) with GAN\end{tabular}\\
\hline
\cite{shokri2017membership} & \begin{tabular}[c]{@{}l@{}}CIFAR10 \cite{krizhevsky2009learning}, \\ CIFAR100 \cite{krizhevsky2009learning}, \\ Purchases \cite{Purchase}, \\ Foursquare \cite{yang2016participatory}, \\ Texas hospital stays \cite{TexaHospital}, \\ MNIST \cite{deng2012mnist}, \\ Adult (income) \cite{UCI}\end{tabular} & \begin{tabular}[c]{@{}l@{}}6,000 images in 10 classes,\\ 60,000 images in 100 classes,\\ 10,000 records with 600 features,\\ 1,600 records with 446 features,\\ 10,000 records with 6170 features,\\ 10,000 handwritten digit images,\\ 10,000 records with 14 attribute\end{tabular} & \begin{tabular}[c]{@{}l@{}}Regarded shadow model\\ resulted as features and\\ label records as in/out\end{tabular} & NN\\
\hline
\cite{salem2019ml} & \begin{tabular}[c]{@{}l@{}}Include 6 sets in \cite{shokri2017membership},\\News \cite{newsgroup},\\Face \cite{learned2016labeled} \end{tabular} & \begin{tabular}[c]{@{}l@{}}Same as above cell,\\ 20,000 newsgroup documents in 20 classes,\\ 13,000 faces from 1,680 individuals\end{tabular} & \begin{tabular}[c]{@{}l@{}}Regarded shadow model\\ resulted as features and\\ label records as in/out\end{tabular} & \begin{tabular}[c]{@{}l@{}}Random Forest,\\ Logistic Regression,\\ Multilayer perceptron\end{tabular}\\
\hline
\cite{ganju2018property}& \begin{tabular}[c]{@{}l@{}}Adult (income) \cite{UCI},\\ MNIST \cite{lecun1998mnist}, \\ CelebFaces Attributes \cite{liu2015faceattributes},\\ Hardware Performance Counters \end{tabular} & \begin{tabular}[c]{@{}l@{}}299,285 records with 41 features,\\ 70,000 handwritten digit images,\\ more than 200K celebrity images,\\36,000 records with 22 features \end{tabular} & \begin{tabular}[c]{@{}l@{}}Neuron sorting ,\\Set-based representation\end{tabular}  & NN \\
\hline
\cite{melis2019exploiting} & \begin{tabular}[c]{@{}l@{}}Face \cite{learned2016labeled},\\ FaceScrub \cite{ng2014data},\\PIPA \cite{zhang2015beyond},\\Yelp-health, Yelp-author \cite{yelp},\\FourSquare \cite{yang2016participatory}, CSI corpus \cite{verhoeven2014clips} \end{tabular} & \begin{tabular}[c]{@{}l@{}}13,233 faces from 5,749 individuals,\\ 76,541 faces from 530 individuals,\\60,000 photos of 2,000 individuals,\\17,938 reviews, 16,207 reviews,\\15,548 users in 10 locations, 1,412 reviews\end{tabular} & N/A & \begin{tabular}[c]{@{}l@{}}Logistic regression,\\gradient boosting,\\Random Forests\end{tabular} \\
\hline
\cite{nasr2018machine} & \begin{tabular}[c]{@{}l@{}}CIFAR100 \cite{krizhevsky2009learning},\\ Purchase100 \cite{Purchase}, \\ Texas100 \cite{TexaHospital} \end{tabular} & \begin{tabular}[c]{@{}l@{}}60,000 images in 100 classes,\\ 197,324 records with 600 features,\\ 67,330 records with 6,170 features\end{tabular} & \begin{tabular}[c]{@{}l@{}}Regarded shadow model\\ resulted as features and\\ label records as in/out\end{tabular} & NN \\
\hline
\end{tabular}
\end{table*}

%% Model inversion attacks that exploit confidence information and basic countermeasures \cite{fredrikson2015model}
\textit{Model Inversion Attack \& Defense:} The model inversion attack was developed by \cite{fredrikson2015model} via conducting the commercial MLaaS APIs and leveraging confidence information with predictions. Though another model inversion attack proposed in \cite{fredrikson2014privacy} leaked the sensitive information from ML's training set, the attack could not work well under other settings e.g.~the training set has a large number of unknown features. However, the attack proposed in \cite{fredrikson2015model} aimed to be applicable across both white-box setting and black-box setting. For the white-box setting, an adversarial client had a prior knowledge about the description of the model as the APIs allowed. For the black-box setting, the adversary was only allowed to make prediction queries on ML APIs with some feature vectors. Considered as the useful data for the attack, the confidence values were extracted from ML APIs by making prediction queries. The attacks were implemented in two case studies --- inferring features of the training dataset, and recovering the training sample of images. The model inversion attack targets the ML model's training data under both settings.

The first attack was inferring sensitive features of the inputs from a decision tree classifier. BigML \cite{BigML} was used to reveal the decision tree's training and querying routines. With query inputs with different features and the corresponding confidential values, the attacker in \cite{fredrikson2015model} accessed marginal priors for each feature of the training dataset. For the black-box setting, the attacker utilized the inversion algorithm \cite{fredrikson2014privacy} to recover the target's sensitive feature with weighted probability estimation. A confusion matrix was used to assess their attacks. For the white-box setting, the white-box with counts (WBWC) estimator was used to guess the feature values. Evaluated with GSS dataset, the results showed that white-box inversion attack on decision tree classifier achieved 100\% precision, while the black-box attack achieved 38.8\% precision. Additionally, the attack in the white-box setting received 32\% less recall than that in the black-box setting. Comparing to black-box attacks, white-box inversion attacks show a significant advance on feature leakage, especially in precision.

The second attack was recovering the images from an NN model from a facial recognition service accessed by APIs. Learning the training samples was required to steal the recognition model firstly. Two specific model inversion attacks were proposed in \cite{fredrikson2015model}, including reconstructing victim's image with a given label, and determining whether the blurred image existed in training set. Specifically, the MI-Face method and Process-DAE algorithm were used to perform the attacks. Herein, the query inputs and confidential values were used to refine the image. The best reconstruction performance from evaluation was 75\% overall accuracy and 87\% identification rate. Moreover, the attacker employed an algorithm named maximum a posterior to estimate the effectiveness. The evaluation results showed that the proposed attacks enhanced the inversion attack efficacy significantly comparing to the previous attack \cite{fredrikson2014privacy}. The training images were recovered accurately.

%% Deep Models Under the GAN: Information Leakage from Collaborative Deep Learning \cite{hitaj2017deep}
\textit{Stealing the Training Data of Deep Model with GAN:} An attack against the privacy-preserving collaborative deep learning was designed to leak the participants' training data which might be confidential \cite{hitaj2017deep}. A distributed, federated, or decentralized deep learning algorithm can process each users' training set by sharing the subset of parameters obfuscated with differential privacy \cite{shokri2015privacy,abadi2016deep}. However, the training dataset leakage problem had not been solved by using the collaborative deep learning model \cite{shokri2015privacy}. An adversary can deceive the model with incorrect training sample to deduce other participants to leak more local data.
%The parameters were generated by each user participated in collaborative deep learning's training phase. 
Then by leveraging the learning process nature, the adversary can train a Generative Adversarial Network (GAN) for stealing others' training samples. The GAN attack targets the collaborative deep learning.

Specifically, the GAN simulated the original model in collaborative learning process to leak the targeted training records \cite{hitaj2017deep}. During the reconnaissance phase, an adversary pretended as one of the honest participants in collaborative deep learning, so that the adversary could influence the learning process and induce the victims to release more information about the targeted class. To collect the valuable data, the adversary did not need to compromise the central parameter server instead of inferring the meaningful information of that class based on the victim's changed parameters. In addition, as a participant was building up the targeted model, part of the training samples were known to the adversary. The fake training dataset could be sampled randomly from other datasets. The true training dataset and a fake training dataset were collected to train the discriminator of the GAN using the CNN algorithm. The outputs of this discriminator and another fake training dataset were used to train the generator of the GAN using CNN. Since the feature of the training data was known by default, the adversary sampled the targeted training data with the targeted label and random feature values. This fake sample was fed into the generator model. The adversary modified the feature values of this fake sample until the predicted label was the targeted label. The final modification of this fake sample was regarded as the target training sample. In the experiments, the GAN attack against collaborative learning was evaluated with MNIST \cite{lecun1998mnist} and AT\&T datasets \cite{samaria1994parameterisation} as inputs. Comparing with model inversion attack, the discriminator within GAN attack reached 97\% accuracy and recovered the MNIST image trained in the collaborated CNN clearly. In a word, the GAN attack trained the discriminator and generator to steal the training data. 

%% Membership Inference Attacks Against Machine Learning Models \cite{shokri2017membership}
\textit{Membership Inference Attack:} Learning a specific data record was targeted by \cite{shokri2017membership}, which was the membership of the training set of the targeted MLaaS model. Since the commercial ML model only allowed black-box access provided by Google and Amazon, not only the training data but also the training data's underlying distribution were controlled. Though the training set and corresponding model were unknown, the output based on a given input revealed the model's behavior. By analyzing such behaviors, adversaries found that the ML model behaved differently on the input that they trained compared to the input which was new to the model. Therefore, according to this observation, an attack model was trained, which could recognize such differences and determine whether the input data was the member of targeted training set or not. The attack is intended to recognize the model's behavior testing with target training sample.

The attack model was constructed by leveraging a shadow training technique \cite{shokri2017membership}. Specifically, multiple ``shadow models'' were built to simulate the targeted model's behavior, which informed the ground truth of membership of their inputs. All ``shadow models'' applied the same service (i.e.~Amazon ML) as the targeted model. In addition, the training data that the adversary used can be generated by the model-based synthesis and statistic-based synthesis methods. The generated dataset shared the similar distribution to the object model's training set, while the testing set was disjoint from training set. Querying these ``shadow models'' with the training sets and testing sets, the prediction results were added a label of $in$ or $out$. These records could be collected as the attack model's training set. Then the adversary utilized the built binary classifier to learn a specific data record by determining whether it was $in$ or $out$ of the training set for MLaaS model. Such an offline attack was difficult to be detected, while MLaaS system would consider the adversary as a legitimate user since the adversary was just querying online. Shadow models were trained to produce the inputs for the membership inference attack.

For this membership inference attack evaluation \cite{shokri2017membership}, several public datasets were used and listed in Table \ref{tab:ML_training}. Three targeted models were constructed by Google Prediction API, Amazon ML, and CNN, respectively. The evaluation metrics used by the adversaries were accuracy, precision, and recall. According to the evaluation results, Google Prediction API was suffered from the biggest training data leakage due to this attack. \skyRe{The accuracy of the attack model was above the baseline 50\% (random guessing result) in all experiments, while the precision were all over 60\%, and recall was close to 100\%.} The membership inference attack learned the training sample effectively.

%For the mitigation, since overfitting was the most important reason that makes ML model be vulnerable to the membership inference attack \cite{shokri2017membership}, regularization techniques could be applied in the ML model to resolve the overfitting problem \cite{srivastava2014dropout, chaudhuri2011differentially, jain2015drop}. Another three mitigation strategies were described as restricting the prediction vector to top $k$ classes, using the coarsen precision results, and increasing entropy of prediction vector for NN models \cite{hinton2015distilling}. The first method, unfortunately, could not fully prevent the membership inference attack. The last two methods obfuscated prediction vectors to mitigate the leaking. Those restriction and obfuscated data protected training set to a limited extent.

%% ML-Leaks: Model and Data Independent Membership Inference Attacks and Defenses on Machine Learning Models \cite{salem2019ml}
In 2019, membership inference attack was further studied in \cite{salem2019ml} to make it broadly applicable at a low cost. Specifically, three assumptions mentioned in \cite{shokri2017membership} are relaxed including using multiple shadow models, synthesizing the dataset from the similar distribution of the target model's training set, and the knowledge of the target model's learning algorithm. The results show that the performance of these attack will not be affected with only one shadow model trained with a dataset from other distributions. The results of using different classification algorithms on one shadow model are not promising. However, by combining a set of ML models trained with various algorithms as one shadow model, the performance of membership inference attack can be tolerable (above 85\% in precision and recall). Herein, the attack is based on an assumption that one model of the model set is trained with the learning algorithm used by the target model. Furthermore, by selecting a threshold of the posterior results to determine the input data's membership, even shadow model is not needed for the membership inference attack. Therefore, the scope of membership inference attack is enlarged.

%% Property Inference Attacks on Fully Connected Neural Networks using Permutation Invariant Representations \cite{ganju2018property}
%% Exploiting Unintended Feature Leakage in Collaborative Learning \cite{melis2019exploiting}
\textit{Property Inference Attack:} Different from learning a specific training record, the property inference attack targets at the properties of training data that the model producer unintended to share.  The target model was defined in \cite{ganju2018property} as a white-box Fully Connected Neural Networks (FCNNs) with the aim to infer some global properties such as a higher proportion of women. To launch this attack and take a model as input, a meta-classifier was built to predict whether the global property exists in this model's training set or not. Above all, several shadow models were trained on a similar dataset using similar training algorithms to mimic the target FCNNs. During the feature engineering phase, the meta-training set was formed in \cite{ganju2018property} by applying set-based representation instead of using a flattened vector of all parameters \cite{ateniese2015hacking}. Specifically, a set-based representation is learned using the DeepSets architecture \cite{zaheer2017deep} in four steps --- 1) flattening each nodes' parameters from all hidden layers, 2) obtaining a node representation with node processing function based on the target property, 3) summing a layer representation with layer summation, and 4) concatenating these layer representations as a classifier representation. The accuracy of this attack reached 85\% or over on binary income prediction, smile prediction or gender classification task. This property inference attack against white-box FCNNs is effective to steal training set information.

In collaborative learning, leaking unintended features about participants' training data is another kind of property inference attack \cite{melis2019exploiting}. Instead of global properties, the unintended feature they targeted is held for a certain subset of training set or even independent of the model's task. For example, the attacker infers black face property of training data while learning a gender classifier in a federated manner. In the reconnaissance process, the adversary as a participant can download the current joint model for each iteration of the collaborative learning. The aggregated gradient updates from all participants are computed, thereafter, the adversary can learn the aggregated updates other than his own updates \cite{mcmahan2017communication}. Since the gradients of one layer are calculated based on this layer's features and the previous layer's error, such aggregated updates can reflect the feature values of other participants' private training set. After several iterations, these updates are labeled with the targeted property and fed to build a batch property classifier. Given model updates as inputs, this classifier can predict corresponding unintended features effectively with most precision values larger than 80\%. Therefore, the collaborative learning is vulnerable to property inference attack as well. 

%% Machine Learning with Membership Privacy using Adversarial Regularization \cite{nasr2018machine}
\sky{\textit{Protection using Adversarial Regularization:} A protection for black-box MLaaS models against the membership inference attack was introduced in \cite{nasr2018machine}.} As described in \cite{shokri2017membership}, membership inference attack could learn whether a data sample was a member of targeted model's training set, even if the adversary only observed the queried output of a cloud service. Regularizing the ML model with L2-norm regularizers was a popular mitigation method \cite{shokri2017membership, fredrikson2015model}, which was not considered to offer a rigorous defense. On the other hand, researchers concluded that differential privacy mechanism prevented this information leakage by sacrificing the model's usability which was expendable. To guarantee the confidentiality and privacy of training set rigorously, there is a need for a privacy mechanism more powerful than regularization and differential privacy.

A defender's objective was analyzed firstly by formalizing membership inference attack in \cite{nasr2018machine}. Precisely, the input of an inference model consisted of a testing data for the targeted classifier, its prediction vector, and a label about membership. An adversary aimed to maximize his inference gain, which was effected by the targeted training dataset and a disjoint dataset for reference attack training. Therefore, the defender intended to minimize the adversary's inference gain while minimizing the loss of targeted classifier's performance. That is, the defender enhanced the security of ML model by training it in an adversarial process. The inference gain as an intermediate result was regarded as the classifier's regularizer to revise the ML model with several training epochs. An adversarial regularization was used in training the classifier.

To evaluate the defense mechanism, three common datasets were used in membership inference attack \cite{shokri2017membership}. The classifier's loss was calculated when the attacker's inference gain reached the highest score \cite{nasr2018machine}. The results showed that the classification loss reduced from 29.7\% to 7.5\% with defense comparing to that without defense for the Texas model, which could be insignificant. For the membership inference attack, the accuracy performance targeted at protected ML model is around 50\%, which was close to the random guessing. 
%That is, membership inference attack no longer helps the adversary to distinguish the training records for targeted ML model. 
In a word, the protection model using the adversarial regularization enhanced the confidentiality and privacy of its training data.

The protection proposed in \cite{nasr2018machine} was powerful against the membership inference attack. However, its effectiveness in protecting training data leaked by other attacks remains unknown. Additionally, it did not discuss whether adversarial regularization can protect the white-box MLaaS models from membership inference attack or not. \sky{Moreover, this defense method failed against the online attack which steals the training data of deep model using a GAN model \cite{hitaj2017deep}.}

%% Semi-supervised Knowledge Transfer for Deep Learning from Private Training Data \cite{papernot2017semi}

\textit{Protection using PATE:} To protect the training set of an ML model generally, Private Aggregation of Teacher Ensembles (PATE) was proposed by \cite{papernot2017semi}. Specifically, PATE prevents training set information leakage from model inversion attack, GAN attack, membership inference attack, and property inference attack. Two kinds of models are trained in this general ML strategy including ``teacher'' and ``student'' models. Teacher models are not published and trained on sensitive data directly. Splitting sensitive dataset into several partitions, several teacher models are trained using learning algorithms independently. These teacher models are deployed as an ensemble making predictions in a black-box manner. Given an input to these teachers, aggregating their predictions as a single prediction depends on each teacher's vote. To avoid that teachers do not have an obvious preference in aggregation, Laplacian noise is added to vote counts. Obtaining a set of public data without ground truth, the student will label them by querying the teacher models. Then the student model can be built in a privacy-preserving manner by transferring the knowledge from teachers. Moreover, its variant PATE-G uses the GAN framework to train the student model with a limited number of labels from teachers. In conclusion, the PATE framework provides a strong privacy guarantee to the model's training set.

%% Pyramid: Enhancing Selectivity in Big Data Protection with Count Featurization \cite{lecuyer2017pyramid}
\textit{Protection using Count Featurization:} A limited-exposure data management system named Pyramid enhanced the protection for organizations' training data storage \cite{lecuyer2017pyramid}. It mitigated the data breaches problem by limiting widely accessible training data, and constructing a selective data protection architecture. For emerging ML workloads, the selective data protection problem was formalized as a training set minimization problem. Minimizing the training set can limit the stolen data.

During the prior data management \cite{tang2012cleanos}, only in-use data were retained in the accessible storage for the ML training periodically. However, the whole dataset would be exposed continuously from updated accessible storage \cite{lecuyer2017pyramid}. For this concern, distinguishing and extracting the necessary data for effective training was the key process. The workflow of Pyramid kept accessible raw data within a small rolling window. The core method named ``count featurization'' was used to minimize the training set. Specifically, the counts summarized the historical aggregated information from the collected data. Then Pyramid trained the ML model with the raw data featurized with counts in a rolling window. The counts were rolled over and infused with differential privacy noise to preserve the training set \cite{dwork2006calibrating}. In addition, the balance between training set minimization and model performances (accuracy and scalability) should also be considered. Three specific techniques were applied to retrofit the count featurization for data protection. The infusion with the weighted noise added less noise to noise-sensitive features of the training set. Another technique, called unbiased private count-median sketch, solved the negative bias problem arising from the noise infusion, while the automatic count selection found out useful features automatically and counted them together. For training data protection, count featurization was used to remain necessary data within data storage. Pyramid prevented the attacker from learning the extracted information from the training set.

\begin{table}[!th] %% Table for ML related info categories 
\scriptsize
\caption{Categories of Stealing ML related information attacks from three perspectives (info: information). As for attack targets, two types of information may be stolen --- model internal information and training set information. From attack surfaces, attacks may occur during either model's training phase or inference phase. Considering the attacker's capability, the ML model usually allows either the black-box access or the white-box access. The first category is used for this subsection's organization.}
\label{tab:ml_cato}
\centering
\begin{tabular}{|l|c|c|c|c|c|c|}
\hline
\multicolumn{1}{|c|}{\multirow{2}{*}{\textbf{Attack Type}}}  & \multicolumn{2}{c|}{\textbf{Attack Targets}}  & \multicolumn{2}{c|}{\textbf{Attack Surfaces}} & \multicolumn{2}{c|}{\textbf{Attacker's Capabilities}}  \\ \cline{2-7} 
\multicolumn{1}{|c|}{}  & Model Info  & Training Set Info & Training Phase  & Inference Phase  & Black-box Access  & White-box Access\\
\hline
Model extraction attack \cite{tramer2016stealing}   & YES & no  & no  & YES & YES & no \\ \hline
Model extraction attack \cite{papernot2017practical}& YES & no  & no  & YES & YES & no \\ \hline
Hyperparameter stealing attack \cite{wang2018stealing} & YES & no  & no  & YES & YES & no \\ \hline
Hyperparameter stealing attack \cite{joon2018towards}  & YES & no  & no  & YES & YES & no \\ \hline
Black-box inversion attack \cite{fredrikson2015model}  & no  & YES & no  & YES & YES & no \\ \hline
White-box inversion attack \cite{fredrikson2015model}  & no  & YES & no  & YES & no  & YES \\ \hline
GAN attack \cite{hitaj2017deep}                        & no  & YES & YES & no  & no  & YES \\ \hline
Membership inference attack \cite{shokri2017membership}& no  & YES & no  & YES & YES & no \\ \hline
Membership inference attack \cite{salem2019ml}      & no  & YES & no  & YES & YES & no \\ \hline
Property inference attack \cite{ganju2018property}  & no  & YES & no  & YES & no  & YES \\ \hline
Property inference attack \cite{melis2019exploiting}& no  & YES & YES & no  & no  & YES \\ \hline
\end{tabular}
\end{table}

\begin{table}[!th] %% Table for Black-box vs White-box attack 
\scriptsize
\caption{Attack's prior knowledge under black-box access and white-box access. The black-box access allows the users to query the model and obtain prediction outputs which include the predicted label and confidence value. The white-box access allows the users to access any information of its model which includes \sky{predicted label, predicted confidence}, parameters, and hyperparameters.}
\label{tab:ml_black_white}
\centering
\begin{tabular}{|l|c|c|}
\hline
\sky{\textbf{Model's Information}} & \sky{\textbf{Black-box Access}} & \sky{\textbf{White-box Access}} \\ \hline
\sky{Predicted Label}    & \sky{YES}   & \sky{YES} \\ \hline
\sky{Predicted Confidence} & \sky{YES}   & \sky{YES} \\ \hline
\sky{Parameters} & \sky{NO} & \sky{YES} \\ \hline
\sky{Hyperparameters}  & \sky{NO} & \sky{YES} \\ \hline
\end{tabular}
\end{table}

\subsubsection{\textbf{Summary}}%%%%%--------------- Summary --------------------%%%%% 
In Section~\ref{subsec_ML}, ML-based stealing attacks against model related information target at either model descriptions or model's training data. In addition to this category, as shown in Table~\ref{tab:ml_cato}, the other two ways focus on attacks at training/inference phase and with black-/white-box access \cite{papernot2018sok}. Model extraction attacks \cite{tramer2016stealing,papernot2017practical} and hyperparameter stealing attacks \cite{wang2018stealing,joon2018towards} leak the model's internal information happened at inference phase. Attackers steal model's training data mostly at inference phase, except the GAN attack \cite{hitaj2017deep} and the property inference attack \cite{melis2019exploiting} which happen at training phase of collaborative learning. When attacking during training phase, attackers with white-box access to the model can exploit its internal information. As shown in Table~\ref{tab:ml_black_white}, the white-box access allows attackers to have more prior knowledge than black-box, which results in high performance of the stealing attack \cite{fredrikson2015model}. On the other hand, black-box attacks can be more applicable in the real world. Except \cite{salem2019ml}, most of the attackers in this category under black-box access know the learning algorithm of the target model \cite{tramer2016stealing,wang2018stealing,papernot2017practical, joon2018towards,fredrikson2015model,li2013membership}.

\textit{Countermeasures:} Concerning the ML pipeline, the protection methods will be applied in data preprocessing phase, training phase, and inference phase respectively. Differential privacy noise used in the first phase can build a privacy-preserving training set \cite{lecuyer2017pyramid}. Differential privacy is the most common countermeasures to defend against the stealing attack, however, it alone cannot prevent the GAN attack \cite{hitaj2017deep}. Differential privacy, regularization, dropout, and rounding techniques are popular protections at the training and inference phases. At the training phase, differential privacy on parameters cannot resist the GAN attack \cite{hitaj2017deep}, while rounding parameters is ineffective against hyperparameter stealing attack \cite{wang2018stealing}. Regularization may be effective depending on the targeted algorithm towards hyperparameter stealing attack \cite{wang2018stealing}.

\subsection{Controlled Authentication Information} \label{subsec_auth} %%%%%%%%%%%%%   Authentication   %%%%%%%%%%%% 
\begin{figure*}[!t] %% Attack at authentication information 
\centering
\includegraphics[width=.6\textwidth]{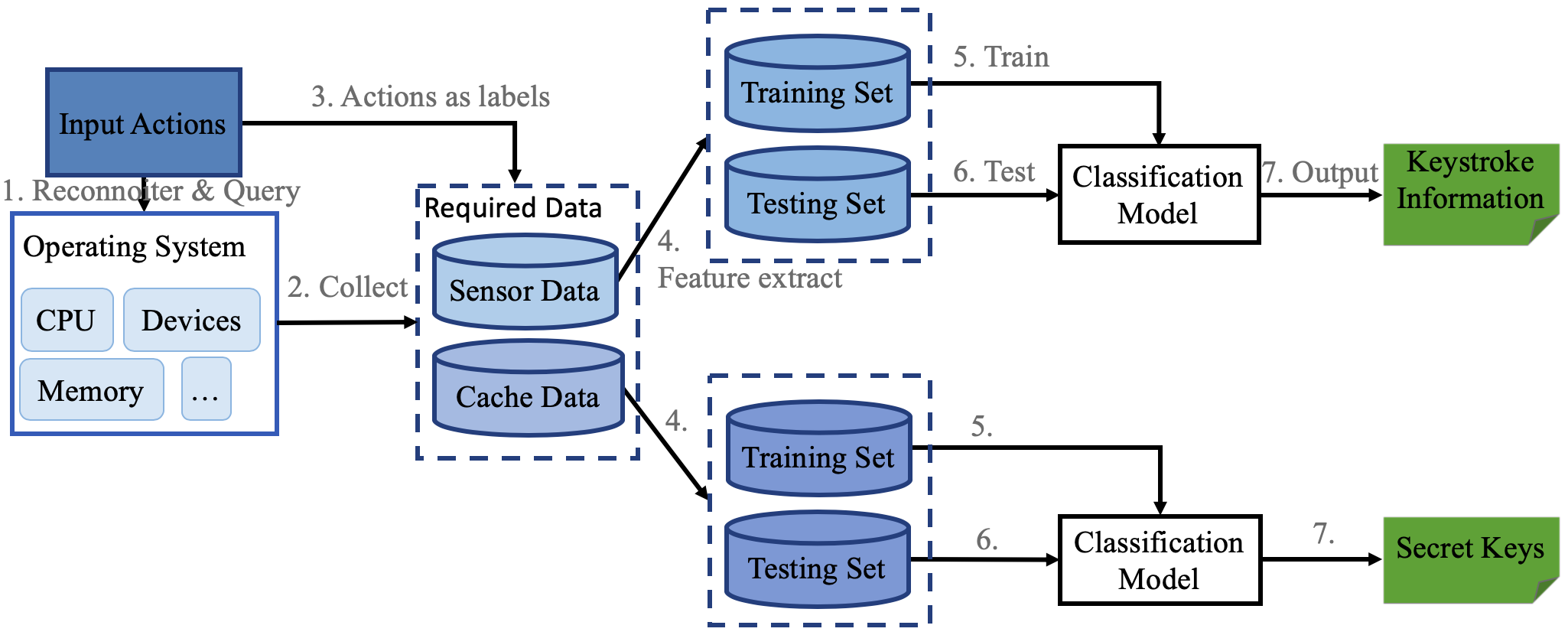}
\caption{The ML-based stealing attack against authentication information --- keystroke information and secret keys. After \textit{reconnoitering and querying}, attackers targeting at keystroke information and secret keys interact with the target system to \textit{collect} data, which refers to the active collection. The attack involved active collection shares a similar workflow as Fig.~\ref{fig:LR_activity} depicted.}
\label{fig:LR_authen_1}
\end{figure*}

The authentication information is one of the most important factors in security while accessing the information from services or mobile applications. Thus, users' authentication information is always stored with protected mechanisms. In Section~\ref{subsec_auth}, the controlled authentication information mainly contains keystroke data, secret keys and password data. As shown in Fig.~\ref{fig:LR_authen_1} and Fig.~\ref{fig:LR_authen_2}, classification models or probabilistic models are trained to steal the controlled authentication information. Additionally, the protections of controlled authentication information stealing attacks are summarized in this subsection.

\subsubsection{\textbf{Stealing controlled keystroke data for authentication}}%%%% Authentication -> keystroke %%%% 
The dataset collected from a device's sensor information can be used to infer the controlled keystroke information as depicted in Fig.~\ref{fig:LR_authen_1}. The keystroke data contains the information about user authentication data, especially for keystroke authentications \cite{alpar2017frequency,krishnamoorthy2018identification,wu2017cost, goodkind2017utilizing}. Leveraging the acceleration, acoustic and video information, we review the attacks stealing these keystroke information and the countermeasures.
\begin{table*}[!th] %% Stealing Controlled Keystroke Data for Authentication %% 
\scriptsize
\caption{Stealing Controlled Keystroke Data for Authentication}
\label{tab:authen_keystroke}
\centering
\begin{tabular}{|c|l|l|l|l|}
\hline
Paper & \multicolumn{1}{c|}{Dataset for Experiment} & \multicolumn{1}{c|}{Description} & \multicolumn{1}{c|}{Feature Engineering} & \multicolumn{1}{c|}{ML-based Attack Method} \\
\hline
\cite{liu2015good} & Acceleration data set & \begin{tabular}[c]{@{}l@{}}Consecutive vectors\\ with 26 labels\end{tabular}\ & \begin{tabular}[c]{@{}l@{}}FFT \& IFFT filter, Movement capturing,\\ Optimization with change direction\end{tabular} & \begin{tabular}[c]{@{}l@{}}Random Forest; \\ $k$-NN; SVM; NN\end{tabular}\\
\hline
\cite{sun2016visible} & Video recordings set & \begin{tabular}[c]{@{}l@{}}Image resolution and\\ frame rate \end{tabular} & \begin{tabular}[c]{@{}l@{}}Extract from selected AOIs' motion \\ signals for motion patterns\end{tabular} & multi-class SVM\\
\hline
\end{tabular}
\end{table*}

%% When Good Becomes Evil: Keystroke Inference with Smartwatch \cite{liu2015good}
\textit{Keystroke Inference Attack:} Several types of sensor information can be utilized to steal keystroke authentication information while targeting at keyboard inputs. In the process of reconnaissance, \cite{liu2015good} found that sensor data from the accelerometer and microphone in a smartwatch was related to user keystrokes. Since the smartwatch was worn on the user's wrist, the accelerometer data reflected the user's hand movement. Therefore, the user's inputs on keyboards can be inferred. The authors presented the corresponding practical attack based on this finding. Adversaries collected the accelerometer and microphone data when inputting the keystrokes in a keyboard. By leveraging the acceleration and acoustic information, adversaries were able to distinguish the content of the typed messages. In addition, two kinds of keyboards were targeted: a numeric keypad of POS terminal and a QWERTY keyboard. The datasets about sensor information were collected for the inference attack.

During the feature engineering phase, adversaries manually defined the $x$-axis and $y$-axis as two movement features of acceleration data, and the frequency features were extracted from acoustic data. Then FFT was employed to filter the linear noise and high-frequency noise. Applying the ML strategies into the attacks, keystroke inference models were set up to reduce the impact caused by the noise within sensor data \cite{liu2015good}. Specifically, the modified $k$-NN algorithm cooperated with an optimization scoring algorithm was applied to enhance the accuracy of their inference attack. Thereafter, the typed information within these two keyboards were leaked, including users' banking PINs and English texts. The attack inferred the keystroke information containing authentication information.

Regarding the evaluation, the results showed that the keystroke inference attack on the numeric keypad had 65\% accuracy in leaking banking PINs among the top 3 candidates \cite{liu2015good}. Unlike the previous work in decoding PINs \cite{cai2011touchlogger,xu2012taplogger, miluzzo2012tapprints}, any devices containing the POS terminal could be compromised by this attack. For the attack targeted at QWERTY keypads, comparing to previous work \cite{berger2006dictionary,backes2008compromising}, a notable improvement had to be achieved to find the word correctly, where the accuracy improved by 50\% with strong allowance to acoustic noise. In the end, several mitigation solutions against the keystroke inference attack were provided \cite{liu2015good}: restricting the access to accelerometer data; limiting the acoustic emanation; and adding the permissions in accessing the sensors which should be managed dynamically according to the context. The attack inferred the keystroke information accurately and was mitigated with the restrictions.

%% VISIBLE: Video-Assisted Keystroke Inference from Tablet Backside Motion \cite{sun2016visible}
\textit{Video-Assisted Keystroke Inference Attack:} Apart from accelerometer and microphone, the video records are another kind of sensor information for attackers to infer the keystroke authentication information. An attack named VISIBLE, provided by \cite{sun2016visible}, leaked the user's typed inputs leveraging the stealthy video recording the backside of a tablet. The attack scenario assumed that the targeted tablet was placed on a tablet holder, and another two types of soft keyboards were used for inputs including alphabetical and PIN keyboards. The dataset for the ML-based attack contained the video of the backside motion of a tablet during the text typing process. In the process of feature engineering, the area of interests (AOIs) were selected and decomposed. The tablet motion was analyzed with amplitude quantified. Then, the features were extracted from temporal and spatial domains. As an ML-based attack, a multi-class SVM was applied to classify various motion patterns to infer the input keystrokes. VISIBLE exploited the relationship between a dictionary and linguistic to refine such inference results. The experiments showed that the accuracy of VISIBLE in leaking input single keys, words and sentences, outperformed the random guess significantly. Particularly, the average accuracy scores of the aforementioned inference attacks were above 80\% for the alphabetical keyboard and 68\% for the PIN keyboard. The countermeasures of this attack include providing no useful information for the video camera, randomizing the keyboards layout, and adding noise when accessing the video camera. The attack leveraged the video information can also infer keystroke authentication information very accurately.

\subsubsection{\textbf{Stealing controlled secret keys for authentication}}%%%%% User Activities -> cache data %%%%% 
Secret keys are used to encrypt and decrypt sensitive messages \cite{chen2016dual,koppula2016deterministic,yan2017protect}. Reconstructing the cryptographic keys means that one host is authenticated to read the message in some cases \cite{paul2015wireless, he2017anonymous, jiang2017lightweight}. However, an adversary has the ability to deduce the sensitive information like cryptographic keys, by understanding the changes in the state of shared caches \cite{gras2018translation}. In this part, the attack stealing controlled secret key information is surveyed via analyzing the state of targeted cache set.
\begin{table*}[!th] %% Stealing Controlled User Activities using Cache Data %% 
\scriptsize
\caption{Stealing Controlled Secret Keys for Authentication (Information: info)}
\label{tab:authen_key}
\centering
\begin{tabular}{|c|l|l|l|l|}
\hline
\textbf{Reference} & \textbf{Dataset for Experiment} & \textbf{Description} & \textbf{Feature Engineering} & \textbf{ML-based Attack Method} \\
\hline
\cite{gras2018translation} & 300 observed TLB latencies & Collect from TLB signals & \begin{tabular}[c]{@{}l@{}}Encode info using a\\ normalized latencies vector\end{tabular} & SVM classifier\\
\hline
\cite{zhou2016software} & 500,000 Prime-Probe trials & \begin{tabular}[c]{@{}l@{}}Number of absent cache\\ lines + cache lines available\end{tabular} & N/A & NB classifier\\
\hline
\end{tabular}
\end{table*}

%% Translation Leak-aside Buffer: Defeating Cache Side-channel Protections with TLB Attacks \cite{gras2018translation}
\textit{Stealing secret keys with TLB Cache Data:} Due to the abuse of hardware translation look-aside buffers (TLBs), secret key information can be revealed by the adversary via analyzing the TLB information \cite{gras2018translation}. The targeted fine-grained information of user memory activities (i.e.~cryptographic keys) was safeguarded in the controlled channels like cache side channels \cite{liu2016catalyst,gruss2017strong}. During the reconnaissance, the legitimate user accesses the shared TLBs, which reflects victims' fine-grained cache memories. In detail, the victim's TLB records could be accessed by other users using CPU affinity system calls or leveraging the same virtual machine. Adversaries reverse-engineered unknown addressing functions which mapped virtual addresses to different TLB sets, in order to clarify the CPU activities from the TLB functions. To design the data collection, the adversary monitored the states of shared TLB sets indicating the functions missed or performed by the victims. Without privileged access to properties of TLB information (i.e.~TLB shootdown interrupts), adversaries timed the accesses to the TLB set and measured the memory access latency, which indicated the state of a TLB set. Instructing the targeted activities with a set of functions statements, the adversary accessed the TLB data shared by the victim and collected the corresponding temporal information as a training dataset. The label, in this case, was the state of the function written in the statement. Datasets about the TLB state information were prepared for the stealing attack.

For the feature engineering, features were extracted from TLB temporal signals by encoding information with a vector of normalized latencies. Additionally, ML algorithms were adopted to distinguish the targeted TLB set by analyzing memory activity. Specifically, with high-resolution temporal features extracted to present the activity, an SVM classifier was built to distinguish the access to the targeted set and other arbitrary sets. In the experiment of \cite{gras2018translation}, the training set contained 2,928 TLB latency data in three different sets. The end-to-end TLBleed attack on \texttt{libgcrypt} captured the changes of the target TLB set, extracted the feature signatures and reconstructed the private keys. During the evaluation phase, TLBleed reconstructed the private key at an average success rate of 97\%. Particularly, a 256-bit EdDSA secret key was leaked with TLBleed successfully with the success rate of 98\%, while 92\% in reconstructing RSA keys. Potential mitigations against the TLBleed attack was discussed in \cite{gras2018translation} including executing sensitive process in isolation on a core, partitioning TLB sets among distrusting processes, and extending hardware transactional memory features. Hence, secret cryptography keys can be reconstructed by distinguishing the targeted TLB set.

%% A software approach to defeating side channels in last-level caches \cite{zhou2016software}
\textit{Protection Against Leakage from CPU Cache Data:} Since an attacker can deduce the private keys from the changes of memory activities, a memory management system should be implemented to prevent these unexpected leakage, like the CacheBar system proposed by \cite{zhou2016software}. In addition to the TLB cache data, this kind of attack can leverage last-level caches (LLCs) \cite{irazoqui2015s,liu2015last} to steal the fine-grained information from victims like secret keys. Preventing the attacker analyzing some valuable information from LLCs, CacheBar dynamically managed the shared memory pages with a copy-on-access mechanism. For example, each domain kept its own copy of one accessed physical page. Another method used in CacheBar was limiting the cacheability of memory pages. For example, only a limited number of lines per cache set could be accessed by attackers. CacheBar was evaluated empirically with a particular attack scenario named the Prime-Probe attack \cite{irazoqui2015s,liu2015last, osvik2006cache}. CacheBar protected the secret key by restricting the access to the CPU cache data.

A Prime-Probe attack with ML strategies was implemented in \cite{zhou2016software}. This attack was conducted when the attacker and the victim ran the programs sharing the same CPU cache sets. In the process of reconnaissance phase, the changes of LLCs were useful for the attack. Targeting at these shared CPU cache sets, the attacker firstly primed the cache sets using its memory. When the victim executed the program loaded in the memory, the conflicts would occur resulting in the absent of some cache set memory. If the attacker probed the cache by calculating the time to load memory into the sharing cache sets, he was able to infer the number of cache lines that were absent in each cache set. To collect training and testing datasets, the attacker run the Prime-Probe attacks several times and recorded the information about the absent cache lines numbers within each cache set. Six classes of these dataset were defined according to the number of absent cache lines \cite{zhou2016software}. For instances, the record with zero absent cache lines was labeled as NONE. An NB classifier was used to determine the classes of absent cache lines, which revealed the changes of different types of cache sets caused by the user. To infer private keys, this ML-based Prime-Probe attack recognized the state of targeted CPU cache set.

For the effect of the protection system, CacheBar significantly mitigated the Prime-Probe attack \cite{zhou2016software}. CacheBar was designed to maintain the queue of cacheable memory pages dynamically. Fundamentally, it limited the number of lines per sharing cache set probed by the attacker, since each physical memory page contained numerous memory blocks mapping to different cache sets. In order to limit the number of cache lines, CacheBar limited the number of domain-cacheable pages whose contents could be loaded in the same cache set. CacheBar also applied one reserved bit within the domain mapping to these non-domain-cacheable pages. Thereafter, the access to any of these pages would be processed by the page fault handler. As a result, by applying CacheBar, the overall accuracy of Prime-Probe attack was decreased significantly from 67.3\% to 33\%. Limiting the access to chacheable CPU caches moderated the threat of the ML-based Prime-Probe attack. 

\begin{figure*}[!t] %% Attack at authentication information 
\centering
\includegraphics[width=.6\textwidth]{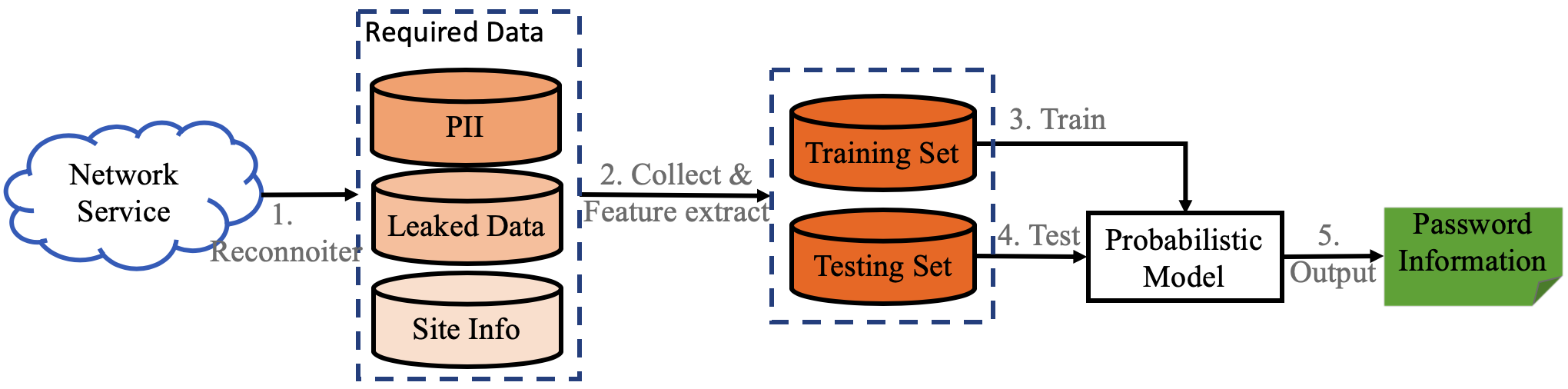}
\caption{The ML-based stealing attack against authentication information --- password data. To infer the password, attackers \textit{reconnoiter and collect} the online information with the passive collection. For the attack with passive collection, attackers do not need to interact the target service with designed inputs. They collect the required data labeled with semantic categories according to human behaviors of password creation \cite{wang2016targeted} or passwords' generic structure \cite{veras2014semantic}. During the \textit{feature engineering} phase, different segments from the required data are extracted. A semantic classifier is \textit{trained} using probabilistic algorithms. After \textit{testing} this classifier, various passwords can be constructed as \textit{outputs} with the semantic generalization.}
\label{fig:LR_authen_2}
\end{figure*} 
\subsubsection{\textbf{Stealing controlled password data for authentication}}%%%%% Authentication -> password %%%%% 
Passwords are considered as one of the most important sensitive information of the user, and its leakage can raise a serious security concern. Most of the useful information to the stealing attack is collected passively from the network services as illustrated in Fig.~\ref{fig:LR_authen_2}. The password guessing attack was studied by analyzing the password patterns with ML techniques.
\begin{table*}[!th] %% Stealing Controlled Password Data for Authentication %% 
\scriptsize
\caption{Stealing Controlled Password Data for Authentication}
\label{tab:authen_psw}
\centering
\begin{tabular}{|c|l|l|l|l|}
\hline
\textbf{Reference} & \textbf{Dataset for Experiment} & \textbf{Description} & \textbf{Feature Engineering} & \textbf{ML-based Attack Method} \\
\hline
\cite{wang2016targeted} & \begin{tabular}[c]{@{}l@{}}Dodonew, CSDN, \\ 126, Rockyou,\\ 000webhost, Yahoo, \\ 12306,\\ Rootkit; \\ Hotel, 51job\end{tabular} & \begin{tabular}[c]{@{}l@{}}16,258,891 (6,428,277) leaked passwords,\\ 6,392,568 (32,581,870) leaked passwords,\\ 15,251,073 (442,834) leaked passwords, \\ 6,392,568 leaked passwords + 129,303 PII,\\69,418 leaked passwords + 69,324 PII;\\ 20,051,426 PII, 2,327,571 PII\end{tabular} & N/A & \begin{tabular}[c]{@{}l@{}}PCFG-based \\algorithm \cite{weir2009password}, \\Markov-based \\algorithm \cite{ma2014study},\\LD algorithm\end{tabular}\\
\hline
\cite{veras2014semantic} & RockYou & 32,581,870 leaked passwords & Segmented with NLP & \begin{tabular}[c]{@{}l@{}}PCFG-based algorithm\end{tabular}\\
\hline
\cite{melicher2016fast} & \begin{tabular}[c]{@{}l@{}}PGS training set \cite{ur2015measuring},\\1class8, 1class16 \cite{kelley2012guess},\\ 3class12 \cite{shay2014can}, 4class8 \cite{mazurek2013measuring},\\ webhost \cite{webhost} \end{tabular} & \begin{tabular}[c]{@{}l@{}}33 million passwords,\\3,062 (2,054) leaked passwords,\\ 990 (990) leaked passwords,\\ 30,000 leaked passwords\end{tabular} & N/A & \begin{tabular}[c]{@{}l@{}}PCFG-based \\algorithm \cite{komanduri2016modeling}, \\Markov models \cite{ma2014study},\\NN\end{tabular}\\ 
\hline
\end{tabular}
\end{table*}

%% Targeted Online Password Guessing: An Underestimated Threat \cite{wang2016targeted}
\textit{Online Password Guessing Attack:} Online password guessing problem and a framework named TarGuess to model targeted online guessing scenarios systematically were introduced by \cite{wang2016targeted}. Since attackers perform an online password guessing attack based on the victim's personal information, systematically summarizing the all possible attack scenarios helps analysts understand the security threats. The architecture of TarGuess was demonstrated with three phases including the preparing phase to determine the targeted victim and build up its password profile, the training phase to generate the guessing model, and the guessing phase to perform the guessing attack. During the reconnaissance phase, according to the diversity of people's password choices, three kinds of information were beneficial in online guessing attacks: PII like name and birthday, site information like service type, and leaked password information like sister passwords and popular passwords. In particular, PII could be divided into two types, including Type-1 PII used to build part of the passwords (i.e.~birthday), and Type-2 PII reflected user behavior in setting passwords (e.g.~language \cite{mazurek2013measuring}). Some leaked passwords were reused by the user. During the data collection phase, the datasets were combined by multiple types of PIIs and leaked passwords. To be more specific, the four TarGuess were listed as TarGuess-I based on Type-1 PII, TarGuess-II based on leaked passwords, TarGuess III based on leaked passwords and Type-1 PII, and TarGuess-IV based on leaked passwords and PII. Datasets for password guessing attacks were prepared.

After dataset was collected with its initial features, attackers adopted probabilistic guessing algorithms including PCFG, Markov n-gram and Bayesian theory to train the four TarGuess models to infer passwords \cite{wang2016targeted}. The accuracy of these four guessing algorithms was evaluated when the guessing time was limited online. Comparing with \cite{li2016study}, the TarGuess-I algorithm outperformed with 37.11\% to 73.33\% more passwords successfully inferred within $10-10^{3}$ guesses. It also outperformed the three trawling online guessing algorithms \cite{bonneau2012science,ma2014study,weir2009password} significantly by cracking at least 412\% to 740\% more passwords. Comparing to \cite{das2014tangled} with 8.98\% success rate, TarGuess-II achieved 20.19\% within 100 guesses. As for TarGuess-III, no prior research could be compared and it achieved 23.48\% success rate within 100 guesses. Concerning TarGuess-IV, the improvements of accuracy were between 4.38\% to 18.19\% comparing to TarGuess-III. Through modeling guessing attack scenarios, a serious security concern was revealed in \cite{wang2016targeted} about online password leakage with effective guessing algorithms. 

%% On the semantic patterns of passwords and their security impact \cite{veras2014semantic}
\textit{Password Guessing with Semantic Pattern Analysis:} An attempt was made in \cite{wang2016targeted} to formalize several passwords guess lists for one targeted user. And a similar attempt was made in \cite{veras2014semantic} to find a general password pattern. A framework presented by \cite{veras2014semantic} built up semantic patterns of passwords for users in order to understand their password security. The security impacts of user's preferences in password creation were identified. For a better reconnaissance, the passwords were analyzed by breaking into two conceptually consistent parts containing semantic and syntactic patterns. Since a password consists of the combination of word and/or gap segments, the attacker intended to understand these patterns by inferring the password's meanings and syntactic functions. By comprehending how well the semantic pattern characterizing the password, plenty of password guesses could be learned for attacks. When those attacks were successful with some guesses, the true passwords were learned. The attack formalized the password with semantic and syntactic patterns.

The password datasets could be collected from password data leakage like the RockYou's password list. Firstly, the NLP methods was used for passwords segmentation and semantic classification. The segmentation was the fundamental step to process the passwords in various forms. The source corpora was a collection of raw words as the segmentation candidates, whereas the reference corpora contained part-of-speech (POS). Specifically, the POS was tagged with Natural Language Toolkit (NLTK) \cite{raj2009resource} based on Contemporary Corpus of American English. With N-gram probabilities representing the frequency of use, the tagged POS was used to select the most suitable segmentation for passwords. After processing the password dataset, the NLP algorithm was used to classify the segments of input passwords and result in a semantic category. Secondly, a semantic guess generator could be built with the PCFG algorithm. Since syntactic functions of the password were structural relationships among semantic classes, the PCFG algorithm was employed to model the password's syntactic and semantic patterns. In detail, this model learned the password grammar from the dataset, generated the guessing sentence of a language \cite{manning1999foundations} with different constructs, and encoded the probabilities of these constructs as output. To learn any true passwords, the semantic guess generator sorted the outputs according to the probability of the password cracking attack. The generator generated a guessing list based on the semantic and syntactic patterns.

To assess the advantage of the semantic guess generator, the success rate of this generator was compared to the result of previous offline guessing attack approach, i.e.~Weir approach \cite{weir2009password}. To crack a password within 3 billion guessing times, 67\% more passwords were cracked by the semantic approach than the Weir approach in terms of LinkedIn leakage \cite{veras2014semantic}. Exploiting the leakage from MySpace, this approach outperformed the Weir approach by inferring 32\% more passwords. %For further discussion, three improvements could be conducted. For example, the attacker could improve the semantic base guessing by considering the unseen words and estimating their probabilities. From the security aspect, the proactive password checker could be set up to use PCFG in order to alert the user and reject the weak passwords. Additionally, the anthropological analysis should be investigated for users periodically and alert users to change its password. Within the limited guessing attempts, the generator correctly guessed the password with a high success rate.

%% Fast, Lean, and Accurate: Modeling Password Guessability Using Neural Networks \cite{melicher2016fast}
\textit{Protection with Modeling Password Guessability:} As a main form of authentication, human-chosen textual password should be resistant to guessing attacks. Evaluating the strength of passwords is one way to avoid creating the password which is vulnerable to the guessing attack. The strength of password can be checked via modeling adversarial password guessing \cite{dell2010password,kelley2012guess,weir2010testing}. According to \cite{melicher2016fast}, it was more successful to apply NNs rather than other password-guessing methods like Markov models \cite{ma2014study, wang2016targeted} and PCFG \cite{veras2014semantic,kelley2012guess,weir2009password}. It was intended to provide a client-side password checker to defend the guessing attack by modeling the passwords guessability using NNs \cite{melicher2016fast}. 

The proactive password checker was implemented to detect weak passwords. The RNN algorithm was applied to guess the text-based password \cite{melicher2016fast}. Firstly, the training datasets were collected from Password Guessability Service (PGS) training set \cite{ur2015measuring}. Instead of processing the password in word-level, the RNN algorithm was used to learn the password text in character-level. Lists of tokens were generated from the training set according to the most 2,000 frequently used tokenized words. During the training phase, the RNN model was trained with the help of transference learning \cite{yosinski2014transferable}, which overcame the sparsity of the training set. Given the preceding character of one password, the guessing model was about to generate the next character till a special password-ending symbol \cite{dell2015monte,ma2014study}. The checker would enumerated all possible passwords when their probabilities exceeded a given threshold. The targeted password's guess number was the number of guesses taken by the attacker to verify the password when all possibilities were in the descending order of likelihood \cite{melicher2016fast}. The lower the guess number was, the higher possibility of the targeted passwords can be guessed. Once the guess number of one created password was lower than a threshold, the checker would alert user to choose another stronger password. All in all, the checker protected the password by alerting its high guessability.

\subsubsection{\textbf{Summary}} %%%%%----- Summary ----%%%%% 
According to different forms of authentications, ML-based stealing attacks target at users' keystroke authentication, secret keys and passwords. As shown in Fig.~\ref{fig:LR_authen_1} and Fig.~\ref{fig:LR_authen_2}, attackers steal users' passwords by cracking the useful information online. For the other two objectives, they exploit the information based on users' activities recorded by an Operating System (i.e.~TLB/CPU cache data). Additionally, password guessing attacks use the probabilistic method to construct a password with the least number of guesses. The attack on the remaining two targets can be transferred as classification tasks by generating keystroke patterns and cache set states.

\textit{Countermeasures:} From the security perspective, two types of countermeasures are introduced as the access restriction and the attack detection. The secret keys, for example, can be protected by managing the accessible related cache data \cite{zhou2016software}. The analysis of password guessability \cite{melicher2016fast} can secure the user's account by setting a strong password. The weak passwords are evaded by detection. The future direction can target the effectiveness of guessing model prediction which is limited by the sparsity of training samples \cite{melicher2016fast}. The defense for the keystroke inference has not been well-developed. The future work may explore the secured access of related sensor data.

\begin{table*}[!th] %%-------------- Summary of Reviewed Papers ---------------%%
\scriptsize
\caption{\skyRe{Summary of reviewed papers from attack, protection, related ML techniques they utilized, and the evaluation metrics.}}
\label{tab:sum_review}
\centering
\begin{tabular}{|c|l|l|l|l|}
\hline
\textbf{Reference} & \textbf{Attack} & \textbf{Protection} & \textbf{Related ML Techniques} & \textbf{Evaluation} \\
\hline
\cite{diao2016no} & \begin{tabular}[c]{@{}l@{}}Unlock pattern \& foreground \\ app inference attack\end{tabular} & \begin{tabular}[c]{@{}l@{}}Restrict access to kernel resources;\\ Decrease the resolution of interrupt data\end{tabular} & \begin{tabular}[c]{@{}l@{}}HMM with Viterbi algorithm;\\ $k$-NN classifier with DTW\end{tabular} & \begin{tabular}[c]{@{}l@{}}Success rate; Time \&\\ battery consumption \end{tabular} \\
\hline
\cite{spreitzer2018procharvester} & Leaking specific events attack & \begin{tabular}[c]{@{}l@{}}Restrict access to kernel resources;\\ App Guardian \cite{AppGuardian,zhang2015leave} \end{tabular} & \begin{tabular}[c]{@{}l@{}}$k$-NN classifier with DTW;\\ Multi-class SVM with DTW\end{tabular} & \begin{tabular}[c]{@{}l@{}}Accuracy;\\ Precision; Recall;\\ Battery consumption\end{tabular} \\
\hline
\cite{xiao2015mitigating} & \begin{tabular}[c]{@{}l@{}}Keystroke timing attack;\\ website inference attack\end{tabular} & \begin{tabular}[c]{@{}l@{}}Design $d^{*}$-private mechanism\end{tabular} & Multi-class SVM classifier & \begin{tabular}[c]{@{}l@{}}Accuracy;\\ Relative AccE\end{tabular} \\
\hline
\cite{zhang2018level} & \begin{tabular}[c]{@{}l@{}}Stealing user activities;\\Stealing in-app activities\end{tabular} & \begin{tabular}[c]{@{}l@{}}Eliminate the attack vectors; Rate limiting;\\ Runtime detection \cite{zhang2015leave}; Coarse-grained\\ return values; Privacy-preserving statistics\\ report \cite{xiao2015mitigating}; Remove the timing channel\end{tabular} & \begin{tabular}[c]{@{}l@{}}SVM classifier;\\ $k$-NN classifier with DTW\end{tabular} & \begin{tabular}[c]{@{}l@{}}Accuracy;\\ Execution time;\\ Power consumption\end{tabular} \\
\hline
\cite{hojjati2016leave} & \begin{tabular}[c]{@{}l@{}}Stealing product's design\end{tabular} & Obfuscate the acoustic emissions & A regression model & Accuracy \\
\hline
\cite{sikder20176thsense} & \begin{tabular}[c]{@{}l@{}}Information leakage via a \\ sensor; Stealing information \\ via a sensor\end{tabular} & \begin{tabular}[c]{@{}l@{}}The contextual model detects malicious\\ behavior of sensors\end{tabular} & \begin{tabular}[c]{@{}l@{}}Markov Chain; NB;\\ Alternative set of ML\\ algorithms (e.g. PART)\end{tabular} & \begin{tabular}[c]{@{}l@{}}Accuracy; FNR;\\ F-measure; FPR;\\Recall; Precision;\\Power consumption\end{tabular} \\
\hline
\cite{tramer2016stealing} & Model extraction attack & \begin{tabular}[c]{@{}l@{}}Rounding confidences \cite{fredrikson2015model};\\ Differential privacy (DP) \cite{dwork2008differential,li2013membership,jagannathan2009practical,vinterbo2012differentially};\\ Ensemble methods \cite{laskov2014practical} \end{tabular} & \begin{tabular}[c]{@{}l@{}}Logistic regression;\\ Decision tree; SVM;\\ Three-layer NN\end{tabular} & \begin{tabular}[c]{@{}l@{}}Test error;\\ Uniform error;\\ Extraction Faccuracy\end{tabular}\\
\hline
\cite{papernot2017practical} & Model extraction attack & \begin{tabular}[c]{@{}l@{}}Gradient masking \cite{goodfellow2015explaining} and defensive \\distillation \cite{papernot2016distillation} for a robust model\end{tabular} & \begin{tabular}[c]{@{}l@{}}DNN; SVM; $k$-NN; Decision\\ Tree; Logistic regression\end{tabular} & Success rate \\
\hline
\cite{wang2018stealing} & \begin{tabular}[c]{@{}l@{}}Hyperparameters stealing attack\end{tabular} & \begin{tabular}[c]{@{}l@{}}Cross entropy and square hinge loss\\ instead of regular hinge loss\end{tabular} & \begin{tabular}[c]{@{}l@{}}Regression algorithms; NN;\\ Logistic regression; SVM\end{tabular} & \begin{tabular}[c]{@{}l@{}}Relative EE; Relative\\ MSE; Relative AccE\end{tabular} \\
\hline
\cite{joon2018towards}&\begin{tabular}[c]{@{}l@{}}Hyperparameters stealing attack\end{tabular} & N/A & Metamodel methods &Accuracy \\
\hline
\cite{fredrikson2015model} & Model inversion attack & \begin{tabular}[c]{@{}l@{}}Incorporate inversion metrics in training;\\ Degrade the quality/precision of the\\ model's gradient information.\end{tabular} & \begin{tabular}[c]{@{}l@{}}Decision Tree;\\ Regression model\end{tabular} & \begin{tabular}[c]{@{}l@{}}Accuracy;\\ Precision;\\ Recall\end{tabular} \\
\hline
\cite{hitaj2017deep} & \begin{tabular}[c]{@{}l@{}}The GAN attack stealing\\ users' training data\end{tabular} & N/A & CNN with GAN & Accuracy \\
\hline
\cite{shokri2017membership} & Membership inference attack & \begin{tabular}[c]{@{}l@{}}Restrict class in the prediction vector;\\Coarsen precision; Increase entropy of the\\ prediction vector \cite{hinton2015distilling}; Regularization\end{tabular} & NN & \begin{tabular}[c]{@{}l@{}}Accuracy;\\Precision;\\ Recall\end{tabular} \\
\hline
\cite{salem2019ml} & Membership inference attack & Dropout; Model Stacking & \begin{tabular}[c]{@{}l@{}}Logistic regression; Random \\Forest; Multilayer perceptron\end{tabular} & \begin{tabular}[c]{@{}l@{}}Precision;\\ Recall; AUC\end{tabular} \\
\hline
\cite{ganju2018property} & Property inference attack & \begin{tabular}[c]{@{}l@{}}Multiply the weights and bias of each\\ neuron; Add noise; Encode arbitrary data\end{tabular} & NN & \begin{tabular}[c]{@{}l@{}}Accuracy;\\Precision; Recall\end{tabular} \\
\hline
\cite{melis2019exploiting} & Property inference attack & \begin{tabular}[c]{@{}l@{}}Share fewer gradients; Reduce input\\ dimension; Dropout; user-level DP\end{tabular} & \begin{tabular}[c]{@{}l@{}}Logistic regression; Gradient \\boosting; Random Forests\end{tabular} & \begin{tabular}[c]{@{}l@{}}Precision;\\ Recall; AUC\end{tabular} \\
\hline
\cite{nasr2018machine} & Membership inference attack & \begin{tabular}[c]{@{}l@{}}Protect with adversarial regularization\end{tabular} & NN & Accuracy \\
\hline
\cite{papernot2017semi} & N/A & \begin{tabular}[c]{@{}l@{}}PATE: transfer knowledge from an\\ ensemble model to a student model\end{tabular} & \begin{tabular}[c]{@{}l@{}}Semi-supervised learning;\\ GAN\end{tabular} & Accuracy \\
\hline
\cite{lecuyer2017pyramid} & N/A & \begin{tabular}[c]{@{}l@{}}Protect stored training data with count-\\based featurization\end{tabular} & \begin{tabular}[c]{@{}l@{}}NN; Gradient boosted tree;\\Logistic regression;\\Linear regression\end{tabular} & \begin{tabular}[c]{@{}l@{}}Average logistic\\squared loss\end{tabular} \\
\hline
\cite{liu2015good} & Keystroke inference attack & \begin{tabular}[c]{@{}l@{}}Restrict access to accelerometer data; Limit\\ acoustic emanation; Dynamic permission\\ management based on context\end{tabular} & \begin{tabular}[c]{@{}l@{}}Random Forest;\\ $k$-NN; SVM; NN\end{tabular} & Success rate \\
\hline
\cite{sun2016visible} & Typed input inference attack & \begin{tabular}[c]{@{}l@{}}Design a featureless cover; Randomize\\ the keyboards' layouts; Add noise\end{tabular} & multi-class SVM & Accuracy \\
\hline
\cite{gras2018translation} & TLBleed attack infers secret keys & Protect in hardware \cite{liu2016catalyst,gruss2017strong} & SVM classifier & \begin{tabular}[c]{@{}l@{}}Success rate\end{tabular} \\
\hline
\cite{zhou2016software} & \begin{tabular}[c]{@{}l@{}}ML-based prime-probe attack\\infers secret keys\end{tabular} & \begin{tabular}[c]{@{}l@{}}CacheBar manage memory pages\\ cacheability\end{tabular} & NB classifier & \begin{tabular}[c]{@{}l@{}}Accuracy; \\Execution time\end{tabular} \\
\hline
\cite{wang2016targeted} & \begin{tabular}[c]{@{}l@{}}Password guessing attack\end{tabular} & N/A & \begin{tabular}[c]{@{}l@{}}PCFG algorithm; Markov \\model; Bayesian theory \end{tabular} & Success rate \\
\hline
\cite{veras2014semantic} & Password guessing attack & N/A & PCFG-based algorithm & Success rate \\
\hline
\cite{zhang2018level} & Password guessing attack & \begin{tabular}[c]{@{}l@{}}Mitigate the threat by modeling password\\ guessability\end{tabular} & \begin{tabular}[c]{@{}l@{}}PCFG-based algorithm;\\ Markov models; NN\end{tabular} & Accuracy \\
\hline
\end{tabular}
\end{table*}

\section{Challenges and Future Works} \label{sec_discussion}
\begin{figure*}[!t] %% challenge of attack & countermeasures %% 
\centering
\includegraphics[width=.6\textwidth]{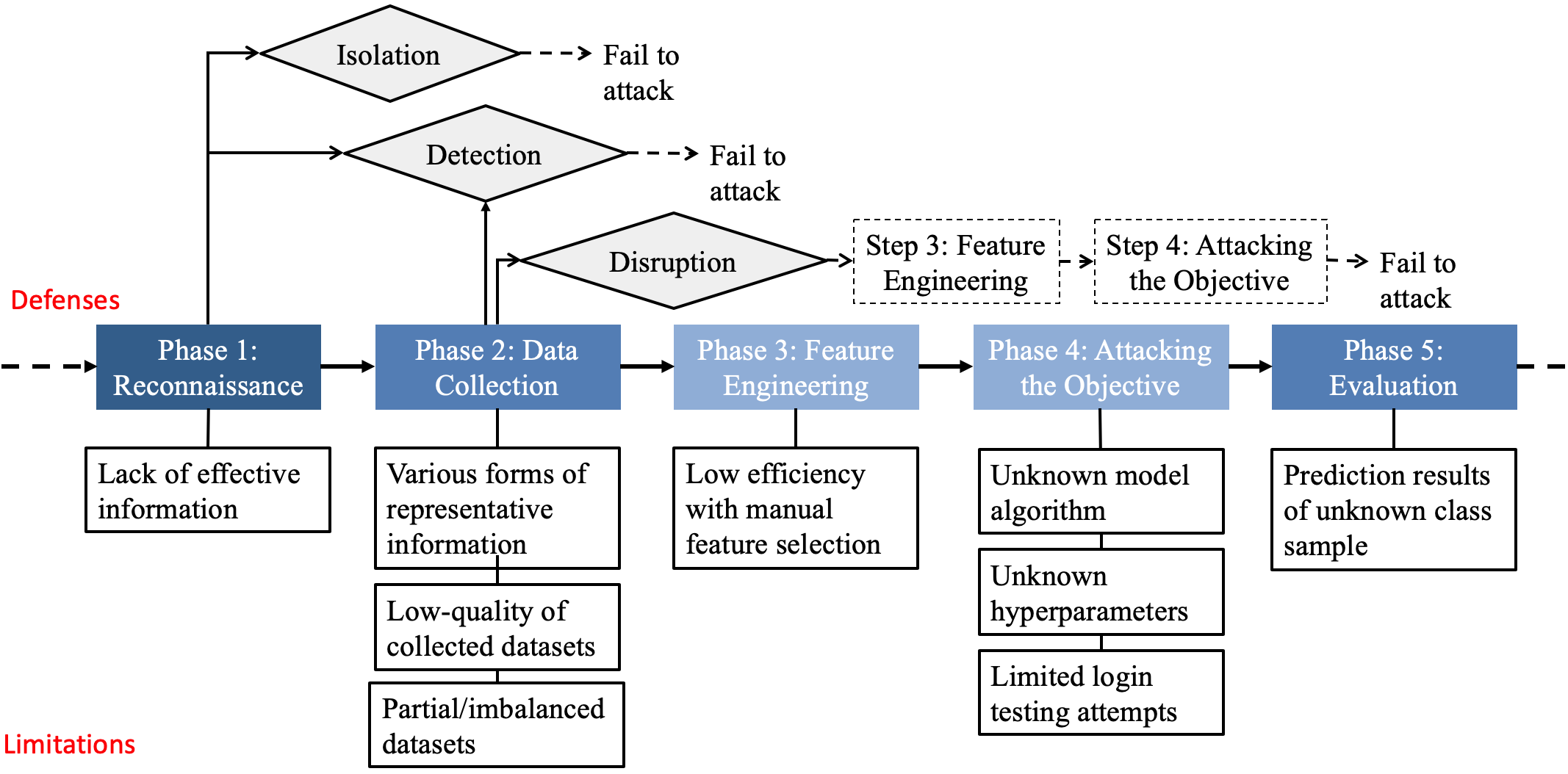}
\caption{The Challenges of ML-based Stealing Attack and Its Defenses} 
\label{fig:Discuss}
\end{figure*}

In Section \ref{sec_LR}, the recent publications about the ML-based stealing attacks against the controlled information and the corresponding defense methods are reviewed. Some attacks can steal the information, but they make strong assumptions of the attacker's prior knowledge. For instance, the attacker is assumed to know the ML algorithm as a necessary condition prior to stealing the model/training samples. However, this prior knowledge is not always publicly known in the real world cases. Additionally, the attack methods are not mature technologies and have great room for improvement. Table~\ref{tab:outline_review} outlines the target and accessible data for each paper. And Table~\ref{tab:sum_review} summarizes the core research papers in the perspectives of attack, protection, related ML techniques, and evaluation. The following subsections will discuss the future directions of the ML-based stealing attack and feasible countermeasures as shown in Fig.~\ref{fig:Discuss}. 

\subsection{Attack} \label{subsec_futureAttack} %% summarize the concepts for each challenge
During the battle between attackers and defenders, it is crucial for defenders to anticipate the strategies of the attackers' future actions. To discuss the future attacks, the challenges of the ML-based stealing attack are analyzed. The analysis results and possible solutions can be summarized and regarded as the future direction of the ML-based stealing attack. In Section~\ref{subsec_futureAttack}, challenges and future directions are discussed from five phases of the MLBSA methodology listed as reconnaissance, data collection, feature engineering, attacking the objective, and evaluation.

\subsubsection{Reconnaissance} %%------------------- Reconnaissance --------------------------- %% 
As illustrated in Section \ref{sec_method_reconnaissance}, the reconnaissance phase consists of two main tasks --- the target definition and valuable accessible data analysis. The denotation of the target determines which kind of accessible resources is valuable. The further attack mechanism is designed according to the analysis of accessible data during the reconnaissance phase. It is essential to ensure that the information accessible to legitimate users contains valuable information for stealing attacks to succeed. 

A challenge of an attack during the reconnaissance phase is the lack of the effective information from the accessible data. As stated in Table \ref{tab:outline_review}, the first category attack --- stealing the user activities information --- primarily relies on the accessible data source including the kernel data and the sensor data. The attacker captures the information without special permissions and utilizes different user activities' representatives as explained in Section \ref{sec_method}. Setting the appropriate permission requirements can protect the accessible data from being exploited by the attacker. For example, Android version 8 restricts the access to kernel resources including interrupt timing log files \cite{spreitzer2018procharvester}. Because of an insufficient amount of information collected under the black-box setting, the model/training data obtained by the stealing attacks is insufficient to reconstruct an ML model as good as the original model \cite{fredrikson2015model, tramer2016stealing, wang2018stealing}. As for the third category of the attack, the majority of the stealing methods are proposed based on a large amount of PII, effective sensor data or coarse-grained cache data. However, these information, especially PII, are sensitive enough to raise privacy concerns and may be protected in the future \cite{shahani2017distributed,chanchary2018privacy}. In summary, current attack vectors restricted to access can block part of ML-based stealing attacks.

Dealing with the lack of information, the future work of the attacker is to find new exploitable data sources as replacements. Some plausible solutions are proposed in \cite{spreitzer2018procharvester, sikder20176thsense}. \cite{spreitzer2018procharvester} defines all interested targets in a list and automatically triggers the activities of interest, followed by searching exploitable logs filed in the newest version of the targeted system (Android version 7 and 8). And it is proven worthless to simply monitor the changes in the accessible data such as the sensor data \cite{spreitzer2018procharvester}, because the detection of the changes gives clues for stealing information. Consequently, future directions can be inspired as searching exploitable source in iOS and monitoring the possible changes of sensor data in order to perform a potential stealing attack. For the attack stealing authentication information, a solution is to use COCA corpus instead of PII. Using the COCA corpus, a successful password guessing attack was performed by \cite{veras2014semantic}. Additionally, analyzing the password structure with anthropological analysis \cite{veras2014semantic} may reduce the attacker's reliance on PII. The attackers were predicted to search new sources or explore new characteristics for further attacks.

\subsubsection{Data Collection} %%-------- Malicious Data Collection -------------------------- %% 
Determining the valuable accessible data is only a part of an ML-based stealing attack. To take advantages of the ML mechanism, the valuable dataset collected in this phase should guarantee its representation, reliability, and comprehensiveness. If either one of three is unsatisfactory, then the results of the stealing attack will be inaccurate. 

The first challenge is collecting valuable data with the representative information of the targeted information including all systems/devices. Especially when the valuable data is kernel data or sensor data, some forms of data recording may vary greatly from systems and devices. Regarding this problem, the data was collected by \cite{diao2016no} and \cite{hojjati2016leave} from eight different mobile devices and different machines. Hence, a future research direction is collecting data from heterogeneous sources and aggregating the representative data. Various forms of representative information affect the attack's probability of success.

The second challenge appears while collecting a reliable dataset. The quality of the training dataset is critical to the attack performance. Most of the explored stealing attacks utilized the model's query output results --- confidential values associated with attacker's query inputs. The preciseness of this value affects the success of the attack. Specifically, the confidential information was leveraged by \cite{tramer2016stealing}, \cite{wang2018stealing} and \cite{fredrikson2015model} to imitate the objective ML model through techniques such as equation solving, path searching, and inversion attack as summarized in Table \ref{tab:ML_description} and Table \ref{tab:ML_training}. Furthermore, the performance of an important attack named membership inference attack \cite{shokri2017membership} depends on the training dataset of the shadow models. The training dataset can be generated based on the target model's confidential information, which is distributed similarly to the targeted training set. Under these circumstances, the aforementioned attacks will not succeed if the target model's API outputs the class only or the polluted confidential information. This inconvenience was scrutinized by \cite{tramer2016stealing}, and a method was proposed to extract the model with only class labels are informed. Accordingly, these findings were further explored in the context of several other ML algorithms together with less monetary advantage when using ML APIs as an honest user. The poor quality of collected dataset hinders the success of the ML-based stealing attack. 

The third challenge of comprehensive dataset collection involves determining the size/distribution of the training dataset and the testing dataset. The size of the training inputs often dictates whether the attacker can easily gain all possible classes of the targeted controlled information, especially when the predictive model outputs only one class per query. In \cite{shokri2017membership}, a comprehensive training dataset was collected by generating the dataset which has a similar distribution to the targets. The size of testing dataset indicates indirectly the amount of controlled information that attackers can learn. For instance, the testing set size of a membership inference model depended on how many training members might be included and would be distinguished \cite{shokri2017membership}. A future work may investigate the impact of the size/distribution of training and testing datasets to the success of ML-based stealing attacks. Partial or imbalanced distribution reduces the success rates of stealing attacks.

\subsubsection{Feature Engineering} %%-------------- Feature Engineering ---------------------- %% 
Feature engineering in the MLBSA methodology intends to refine the collected data for the effective and efficient training process. It is critical to the performance of ML-based attack by eliminating the noise from the collected data. However, among the current research, the techniques used in feature engineering remain underdeveloped.

As shown in Table \ref{tab:user_kernel}, Table \ref{tab:user_sensor}, Table \ref{tab:authen_keystroke}, and Table \ref{tab:authen_psw}, many existing solutions select features manually. Manual feature selection, relying on the attacker's domain-specific knowledge and human intelligence, usually produces a small number of features. That is, manual feature selection is inefficient because of the nature of human involvements, and it may ignore the useful features with low discriminative power. To improve the attack's effectiveness, the automation of feature selection has great research potentials. For example, CNN was used in \cite{hitaj2017deep} to learn a representation of features based on the correlation among data for optimal classification. A future trend of the attack is searching for or developing other automatic methods to overpower the manual feature selection \cite{Coulter2020DDCS}. In \cite{kaul2017autolearn}, a regression-based feature learning algorithm was developed to select and generate features without domain-specific knowledge required. Automating feature selection with such generic algorithm would promote the efficiency and effectiveness of the ML-based attack.

\subsubsection{Attacking the Objective} %%---------- Attacking the Objective: ML algorithm choose -------- %% 
In the phase of attacking the objective with ML techniques, the main tasks include training and testing the ML model to steal the controlled information. There are a few challenges of stealing attacks with respect to training and testing ML models including unknown model algorithms, unknown hyperparameters of ML model, and the limited amount of testing time. 

For ML-based attack stealing the controlled model/training data, the first challenge is that most of the research considered the model algorithm as a prior knowledge. However, the model algorithms for many MLaaS are unknown to the end-user. Most of the attacks would not succeed without specifying the correct model algorithm. In \cite{wang2018stealing}, this concern was discussed, afterwards, a conclusion was drawn that attacks without understanding the model algorithm can be impossible in some circumstances. It is worth investigating the possibility of success for an attack in the context of the unknown model algorithm. In 2019, membership inference attack was considered in \cite{salem2019ml} against a black-box model by choosing a threshold to reveal model's training set information. However, it remains unknown whether this method is applicable to other attacks under the black-box access like the parameter stealing attack \cite{tramer2016stealing}.

The second challenge involves the unknown hyperparameters of the ML model learned from stealing attack comparing to the targeted model. The more precise the model learned, the more accurate the model's functionality; the more precise the model learned, the more detailed the training records that will be revealed. The stealing attack predominantly stole the model by reckoning the parameters of matching objective functions. However, hyperparameter as a critical element  has been ignored, the values of which influence the accuracy of stealing attacks. In \cite{wang2018stealing}, a solution was proposed to prevent an attacker calculating the hyperparameters of some ML algorithms which consist of a set of linear regression algorithms and three-layer neural networks. The future direction toward solving this difficulty can enable a hyperparameter stealing attack against other popular ML algorithms such as $k$-NN and RNN. With unknown hyperparameters, only the parameters can be calculated while extracting the ML model.

The third category of stealing attacks is password guessing attacks. This type of attack generally assumes an unlimited login testing attempts for each account. One exception is in \cite{wang2016targeted} where each login password attack was performed less than 100 times. To crack the password effectively, researchers applied ML algorithms to analyze the password generator based on personal information, website-related information and/or previous public leaked passwords. The future work for this stealing attack can be a successful attack mechanism, which is designed for the targeted authentication system with less than 100 times for login testing. With limited login testing attempts, guessing attack may be failed with the first a few guesses.

\subsubsection{Evaluation} %%----------------------- Evaluation ------------------------------------------ %% 
To effectively infer the controlled information, most of the investigated research applied ML mechanism mentioned in Section \ref{sec_LR}. The prediction of the unknown testing samples is a challenge for ML-based stealing attacks, as the supervised learning algorithm dominates the attack methods. That is, if the true label of a testing sample has not been learned by the model during the training phase, this sample will be recognized as an incorrect class. The testing samples, which are unknown to the training dataset, affect the evaluation results and subsequently reduce the stealing attack's accuracy. To improve the performance of such attacks, the attacker needs to achieve breakthroughs towards predicting the unknown data.

For stealing the user activities attack, when an attacker wants to know the foreground app running in a user's mobile, some distinctive features of the accessible data set which represents the status of running apps will be extracted and learned by ML algorithms \cite{diao2016no, zhang2018level, xiao2015mitigating}. After the attack model is trained, the accessible data recording a new foreground app running in the mobile is the testing sample, this new app is unknown to the attack model \cite{diao2016no, zhang2018level, xiao2015mitigating, li2013membership}. For stealing authentication information attack, these attacks are difficult to be effective when users change their passwords frequently or adopt new layouts for the keyboard \cite{sun2016visible}. This is owing to the uncertainty of the users' password generation behaviors and the variety of users' input keyboards. Evaluating the prediction of unknown class is a challenging task for stealing attack against the user activities and the authentication information.

\subsection{Defense} %%------------------ Countermeasures ------------------------------------- %% 
Targeting diverse controlled information, the countermeasures in protecting the information from ML-based stealing attacks are summarized. In general, the countermeasures can be summarized into three groups: 1) the detection is indicated as detecting related critical indications; 2) the disruption intends to break the accessible data at a tolerable cost of service's utility; and 3) isolation aims to limit the access to some valuable data sources. As shown in Fig.~\ref{fig:Discuss}, the countermeasures mainly applied in the first two phases. Specifically, isolation restricts the attacker's access and makes the attack fail at the first phase; and disruption may confuse the attacker in the second phase and hinder the attacker to build a successful attack model. The detection techniques can detect the attacker's actions and then protect the information from being stolen. These issues are explained as follows. 

\subsubsection{Detection} % no future direction & challenge
To detect potential stealing attacks in advance, the relevant crucial indications are required by analyzing the functionality related to the controlled information. Defenders should notice the attackers' actions as soon as the attackers start the reconnaissance or the data collection processes. Based on the attacker's future directions, the detection is proposed accordingly in order to prevent the attack at an early stage and minimize the loss of stealing the controlled information.

In the presence of any malicious activities in the stage of reconnaissance and data collection, the change or the usage of relevant crucial information should be analyzed and checked. For example, when the attacker steals the information based on the accessible sensor data calling from some APIs, the calling rate and the API usage can be deemed as two critical indications for detection \cite{zhang2018level}. Since attackers may intend to exploit unknown critical indications, a defender can trade off between the access frequency of all related information and the service's utility. Thereafter, the defender detects the unusual access frequency against the stealing attack.

Another detection method is assessing the ability of the service securing the controlled information. This protection can promise that the ML-based attack for stealing information is less powerful than the current attacks. The memory page's cacheability was managed by \cite{zhou2016software} to protect secret keys within user's memory activities, while the password's guessability was checked by \cite{melicher2016fast}. Thereafter, users were alarmed with the weak password. If a defender can assess the ability to hide the ML model and training set from unauthorized peeps, the controlled information are protected to some extent. The detector alerts the user if the assessed ability is below a predefined threshold.

\subsubsection{Disruption} 
Disruption can protect the controlled information via obstructing the information used in each phase of the MLBSA methodology. Disrupting the accessible data currently involves two methods as adding noise to data sources and degrading the quality/precision of service's outputs. For more advanced countermeasures, further research needs to better understand the attacker's future directions.

By disrupting the accessible data, attackers cannot find out valuable accessible sources in the reconnaissance phase, get reliable dataset in the collection phase, or use feature engineering effectively. Therefore, disruption minimizes the success rate of the ML-based stealing attack. The major technique for adding noise is the differential privacy \sky{(DP)} as applied in \cite{xiao2015mitigating,zhang2018level,lecuyer2017pyramid,dwork2008differential, li2013membership}. \sky{Furthermore, DP can be categorized as three categories of disruption methods including global DP \cite{dwork2014algorithmic}, local DP \cite{dwork2014algorithmic} and distributed DP \cite{lyu2018ppfa, lyu2017privacy, lyu2019distributed}. The global DP can be applied on model's global parameters, and local DP and distributed DP are mainly used on collaborative schemes similar to a GAN attack \cite{hitaj2017deep} at the model's parameter-level and record-level. According to \cite{hitaj2017deep}, the record-level DP could be more robust against the GAN attack than the parameter-level DP. From another perspective, the record-level DP does not affect the model stealing attacks \cite{tramer2016stealing, wang2018stealing}, but all kinds of DP can protect training set information at a certain level.} As for the latter method, the specific techniques include rounding the values of outputs (i.e.~coarse-grained predictions and/or confidence) \cite{zhang2018level, tramer2016stealing, fredrikson2015model, shokri2017membership}, and regularizing or obfuscating the accessible data sources \cite{diao2016no, sikder20176thsense, lecuyer2017pyramid, nasr2018machine, sun2016visible}. 

The advanced disruption methods against the attacker's further attacks are also considered. Assuming that attackers may search and collect a class of information from a few devices in the future, advanced disruption should premeditate adding different noise for the similar information shared between various devices. To prevent the advanced feature engineering techniques and ML-based analysis that attackers might apply, complicated and skillful methods for disruption can be applied to defend the controlled information  such as an adversarial training algorithm acting as a strong regularizer \cite{nasr2018machine}. 

\subsubsection{Isolation} 
Isolation can assist the system by eliminating the information stealing threat, which hinders the attacker from progressing through the reconnaissance phase. No matter how attackers improve their strategies and techniques, isolation can protect the controlled information by restricting access to the data of interest. Specifically, it is effective to control the accessible data via restricting the access or managing the dynamic permission \cite{diao2016no, spreitzer2018procharvester, shokri2017membership, liu2015good}. When the stealing attacks advance, defenders can apply ML techniques to automatically control as many as possible accesses related to the targeted controlled information. However, this protection is highlighted to be applied cautiously by concerning the utility of the service. On the one hand, specialists can remove the some information channels which may reveal valuable information to the adversary \cite{zhang2018level, gras2018translation}. On the other hand, if attackers find new exploitable accessible sources in the future, then it is challenging to isolate all the relevant data while ensuring the service's utility. Isolation effectively protects the information by restricting access to the data.

% There is another method for isolation. Specifically, it is effective to control the accessible data via restricting the access or managing the dynamic permission \cite{diao2016no, spreitzer2018procharvester, shokri2017membership, liu2015good}. Concerning attackers may improve their stealing attacks, defenders can apply ML techniques to automatically control all accesses related to the targeted controlled information. 

\section{Conclusion} \label{sec_conclusion}
In this survey, the ML-based stealing attack against the controlled information and the defense mechanisms are reviewed. The generalized MLBSA methodology compatible with the published work is outlined. Specifically, the MLBSA methodology uncovers how adversaries steal the controlled information in five phases, \textit{i.e.}~reconnaissance, data collection, feature engineering, attacking the objective, and evaluation. Based on different types of the controlled information, the literature was reviewed in three categories consisting of the controlled user activities information, the controlled ML model related information, and the controlled authentication information. The attacker is assumed to use the system without any administrative privilege. This assumption implies that user activities information was stolen by leveraging the kernel data and the sensor data both of which are beyond the protection of the application. The attack against the controlled ML model-related information is demonstrated with stealing the model description and/or stealing the training data. Similarly, keystroke data, secret keys, and password data are the examples of stealing the controlled authentication information.

Besides the stealing attack, the corresponding protections are summarized for each category. After reviewing the recent research, it is essential to indicate the challenges clearly in the majority of existing work according to the five attacking phrases.

The future directions matching various limitations are presented. Comparing to the explicit breaking/destroying attack, the controlled information leaked by such stealing attacks is much more difficult to be detected, so that the estimated loss should be extended accordingly. This survey, therefore, can help researchers familiarize these stealing attacks, their future trends, and the potential defense methods.  

\bibliographystyle{ACM-Reference-Format}
\bibliography{ref_leak_ML}

\end{document}